\documentclass[10pt,a4paper]{article}

\usepackage{bm}
\usepackage{bbm}
\usepackage[final]{graphics}
\usepackage{amsmath}
\usepackage{amsfonts,amsbsy}
\usepackage{amssymb}
\usepackage{geometry}
\usepackage{titlesec}
\usepackage{subfigure}
\usepackage{amsmath}
\usepackage{float}
\usepackage{amssymb}
\usepackage{cancel}
\usepackage{xcolor}
\usepackage{slashed}
\definecolor{linkcolor}{rgb}{0.4,0.1,0.1}
\definecolor{bibcolor}{rgb}{0.4,0.1,0.1}
\usepackage[sort&compress,numbers]{natbib}
\usepackage[colorlinks,linkcolor=linkcolor,
citecolor=bibcolor,pdfpagelabels,hyperindex=true]{hyperref}
\def\empile#1\over#2{\mathrel{\mathop{\kern 0pt#1}\limits_{#2}}}
\def\bs{\boldsymbol}

\newcommand{\slv}{\raise.15ex\hbox{$/$}\kern-.53em\hbox{$v$}}
\newcommand{\slF}{\raise.15ex\hbox{$/$}\kern-.53em\hbox{$F$}}
\newcommand{\slL}{\raise.15ex\hbox{$/$}\kern-.53em\hbox{$L$}}
\newcommand{\slP}{\raise.15ex\hbox{$/$}\kern-.53em\hbox{$P$}}
\newcommand{\slp}{\raise.15ex\hbox{$/$}\kern-.53em\hbox{$p$}}
\newcommand{\slq}{\raise.15ex\hbox{$/$}\kern-.53em\hbox{$q$}}
\newcommand{\slR}{\raise.15ex\hbox{$/$}\kern-.53em\hbox{$R$}}
\newcommand{\slQ}{\raise.15ex\hbox{$/$}\kern-.53em\hbox{$Q$}}
\newcommand{\slK}{\raise.15ex\hbox{$/$}\kern-.53em\hbox{$K$}}
\newcommand{\slk}{\raise.15ex\hbox{$/$}\kern-.53em\hbox{$k$}}
\newcommand{\slD}{\raise.15ex\hbox{$/$}\kern-.73em\hbox{$D$}}
\newcommand{\slC}{\raise.15ex\hbox{$/$}\kern-.53em\hbox{$C$}}
\newcommand{\slA}{\raise.15ex\hbox{$/$}\kern-.53em\hbox{$A$}}
\newcommand{\slSigma}{\raise.15ex\hbox{$/$}\kern-.53em\hbox{$\Sigma$}}
\newcommand{\slpartial}{\raise.15ex\hbox{$/$}\kern-.53em\hbox{$\partial$}}
\newcommand{\slcalP}{\raise.15ex\hbox{$/$}\kern-.63em\hbox{$\cal P$}}
\renewcommand{\d}{\ensuremath{\mathrm{d}}}

\def\p{{\boldsymbol p}}

\def\l{{\boldsymbol l}}
\def\k{{\boldsymbol k}}

\def\x{{\boldsymbol x}}

\catcode`\@=11

\newcount\@tempcntc
\def\@citex[#1]#2{\if@filesw\immediate\write\@auxout{\string\citation{#2}}\fi
  \@tempcnta\z@\@tempcntb\m@ne\def\@citea{}\@cite{%
        \@for\@citeb:=#2\do%
    {\@ifundefined{b@\@citeb}%
        {\@citeo\@tempcntb\m@ne\@citea%
                \def\@citea{,\penalty\@m\ }{\bf ?}\@warning%
                {Citation `\@citeb' on page \thepage \space undefined}}%
        {\setbox\z@\hbox{\global\@tempcntc0\csname b@\@citeb\endcsname\relax}
     \ifnum\@tempcntc=\z@ \@citeo\@tempcntb\m@ne%
       \@citea\def\@citea{,\penalty\@m}%
       \hbox{\csname b@\@citeb\endcsname}%
     \else%
      \advance\@tempcntb\@ne%
      \ifnum\@tempcntb=\@tempcntc%
      \else\advance\@tempcntb\m@ne\@citeo%
      \@tempcnta\@tempcntc\@tempcntb\@tempcntc\fi\fi}}\@citeo}{#1}}%

\def\@citeo{\ifnum\@tempcnta>\@tempcntb\else\@citea
  \def\@citea{,\penalty\@m}%
  \ifnum\@tempcnta=\@tempcntb\the\@tempcnta\else
   {\advance\@tempcnta\@ne\ifnum\@tempcnta=\@tempcntb \else
\def\@citea{--}\fi
    \advance\@tempcnta\m@ne\the\@tempcnta\@citea\the\@tempcntb}\fi\fi}

\catcode`\@=12

\begin{document}
\title{\bf %
Instability induced
pressure isotropization\\ in a longitudinally expanding system}
\author{Kevin Dusling${}^{(1)}$, Thomas Epelbaum${}^{(2)}$, Fran\c cois Gelis${}^{(2)}$, Raju Venugopalan${}^{(3)}$}
 \maketitle
 \begin{center}
 \begin{itemize}
 \item[(1)]  Physics Department, North Carolina State University,\\
    Raleigh, NC 27695, USA
 \item[(2)] Institut de Physique Th\'eorique (URA 2306 du CNRS)\\
   CEA/DSM/Saclay, 91191 Gif-sur-Yvette Cedex, France
 \item[(3)]  Physics Department\\
   Brookhaven National Laboratory, Upton, NY-11973, USA
 \end{itemize}
 \end{center}

\begin{abstract} 
  In two previous works [arXiv:1009.4363,arXiv:1107.0668], we 
  studied the time evolution of a system of real scalar fields with
  quartic coupling which shares important features with the Color
  Glass Condensate description of heavy ion collisions. Our primary objective was to understand how such a system, when initialized with a
  non-perturbatively large classical field configuration, reaches
  thermal equilibrium. An essential goal of these works was to
  highlight the role played by the quantum fluctuations.

  However, these studies considered only a system confined within a
  box of fixed volume. In the present paper, we extend this work to a
  system that expands in the longitudinal direction thereby more
  closely mimicking a heavy ion collision. We conclude that the
  microscopic processes that drive the system towards equilibrium are
  able to keep up with the expansion of the system; the pressure
  tensor becomes isotropic despite the anisotropic expansion.
\end{abstract}

\section{Introduction} \label{sec:intro}

The issue of thermalization of the quark-gluon matter produced in
heavy ion collisions is one of the most challenging problems in this
field.  On the one hand, this matter appears to behave like a nearly
perfect fluid, as is suggested by the comparison between flow
measurements at RHIC \cite{Adamsa3,Adcoxa1,Arsena2,Backa2} and
hydrodynamical models with low shear viscosity
\cite{RomatR1}. Moreover, this comparison seems to call for a very
early onset ($< 1~$fm/c) of the hydrodynamical behavior. Such a good
description by hydrodynamics is usually interpreted as an indication
that the system is fairly close to local thermal equilibrium. On
the other hand, trying to justify this fast equilibration from first
principles has proven to be notoriously difficult and no definitive
conclusion has been reached thus far. Early estimates of the
thermalization time based on a standard kinetic description (with
$2\to 2$ and $2\to 3$ processes), as in the bottom-up scenario
\cite{BaierMSS2}, lead to equilibration times that are much larger than what
hydrodynamics seems to require. In addition to kinetic scattering processes, 
quarks and gluons with an anisotropic particle distribution are subject to Weibel instabilities \cite{Mrowc3,Mrowc4} that are well known 
in QED plasmas. These potentially play a significant role in restoring isotropy and in thermalizing the
system. A formal description of Weibel instabilities can be couched in the framework of Hard Thermal Loop effective field theory 
where the hard particles coexist with soft fields
\cite{RebhaRS1,RebhaRS2,MrowcRS1,RomatS1,RomatS2}. This approach has been applied to study
the evolution of such anisotropic systems, and the role of
instabilities in their relaxation towards equilibrium, in a series of
mostly numerical works
\cite{RebhaS1,RebhaSA1,ArnolLM1,ArnolLMY1,ArnolM1,ArnolM2,ArnolM3,ArnolMY4,DumitNS1,BodekR1,BergeGSS1}. For
recent analytical parametric estimates of the effect of Weibel
instabilities in the expanding systems produced in heavy ion collisions,
see refs.~\cite{KurkeM1,KurkeM2}.

Another feature of heavy ion collisions that complicates our understanding of
thermalization is the fact that, at such early times, the 
collision is more appropriately described in terms of fields rather than
on-shell particles\footnote{Soft modes, as we know from the uncertainty principle,  need a rather long time to go
  on-shell.}. Such a description arises naturally in the Color Glass
Condensate (CGC) effective theory. In this approach
\cite{McLerV1,McLerV2,McLerV3,IancuLM3,IancuV1,GelisIJV1,Lappi6}, the nuclei are described in terms of highly boosted color 
sources coupled to soft gauge fields. At the very high energies reached at RHIC and LHC, the density of color sources becomes so large that 
their currents are inversely proportional to the (assumed to be small) coupling constant. In this
framework, one can systematically arrange the various contributions to a computation of inclusive quantities in
powers of the strong fine structure constant $\alpha_s$ with a given order in
$\alpha_s$ corresponding to a fixed loop order.  The leading order (LO)
approximation can be expressed entirely in terms of the classical
solution of the Yang-Mills equations, the next-to-leading (NLO) order
is composed of the 1-loop graphs in the presence of a classical background
field, and so on.  A very satisfying aspect of this description is that the
color source distribution of the projectiles appears only via universal weight functionals of these distributions. Factorization theorems have been proven that demonstrate (to leading accuracy in powers of $\alpha_s Y$, where $Y$ is the rapidity) that these weight functionals are the same for all inclusive observables and for different reactions involving identical projectiles \cite{GelisLV3,GelisLV4,GelisLV5}. The evolution of these weight functionals with rapidity is described by the JIMWLK equation\cite{JalilKMW1,JalilKLW1,JalilKLW2,JalilKLW3,JalilKLW4,IancuLM1,IancuLM2,FerreILM1}.

In the CGC framework, the Weibel instabilities manifest themselves in
the form of unstable solutions of the classical Yang-Mills equations
\cite{BiroGMT1,HeinzHLMM1,RomatV1,RomatV2,RomatV3,FujiiI1,FujiiII1,KunihMOST1,FukusG1},
that lead to secular divergences in the NLO\footnote{At LO, the
  pressure is finite at all times, but the longitudinal pressure is
  negative \cite{KrasnV1,KrasnNV4,LappiM1}.} correction to quantities
such as the pressure: these corrections diverge when the time goes to
infinity. In ref.~\cite{GelisLV2}, we sketched an improvement that
resums, at each order of the expansion in $\alpha_s$, the terms that
have the fastest growth in time. This resummation scheme, which
amounts in practice to averaging classical solutions of the field
equations of motion over a Gaussian ensemble of initial
conditions\footnote{To the best of our knowledge, this scheme was
  first used in \cite{PolarS1,Son1,KhlebT1}, in the context of
  preheating in post-inflationary cosmology. See also ref.~\cite{MichaT1}. In
  the context of non-abelian plasma instabilities, it was used already
  in \cite{BergeSS2}. A similar approach has also been applied to cold
  atom physics, in problems related to Bose-Einstein condensation (see
  for instance refs.~\cite{Norri1,NorriBG1}).}, was justified more
extensively in \cite{FukusGM1,DusliGV1}. Moreover, the precise form of
the initial Gaussian ensemble of gauge fields was also derived in this
work.

Since the numerical implementation of this resummation in QCD is
computationally very demanding, we first considered a much simpler field theory in
\cite{DusliEGV1,EpelbG1}, a scalar field theory with a quartic $\phi^4$ coupling. Similar
issues arise in this simpler context and provide one with a feeling for the effect of
instabilities on the evolution of the system towards equilibrium. The scalar theory, like QCD, is scale invariant at the classical level in $3+1$
dimensions\footnote{This property is not mandatory, but it
  has simple implications for the equation of state: $\epsilon=3P$, up
  to small corrections.}. More importantly, this theory is known to have unstable classical solutions because of parametric resonance
(see refs.~\cite{GreenKLS1,BergeS3}). Although the microscopic nature of these
instabilities is very different from the Weibel instabilities
encountered in Yang-Mills theory, they also lead to secular divergences
in the pressure at NLO.  Therefore, as in the latter,  a resummation of the divergent contributions at each order is essential to 
obtain finite results at late times. 

In refs.~\cite{DusliEGV1,EpelbG1}, a resummation identical to the one
advocated in \cite{GelisLV2,FukusGM1,DusliGV1} for QCD was applied to this
scalar theory and implemented in a numerical lattice computation.
These simulations showed that the pressure quickly relaxes to the
value required by the equilibrium equation of state
($P=\epsilon/3$). On longer timescales, the particle distribution
evolves to a classical equilibrium distribution. Moreover we found
that when the system is initialized with an excess of particle number
(compared to the equilibrium value at a given energy), then a
transient chemical potential appears, and even a Bose-Einstein
condensate\footnote{These two features can only
be transient because the particle number is not conserved in this
theory and eventually inelastic processes will eliminate any
excess in the particle number.} if the particle number excess is too large (see also
refs.~\cite{BlaizGLMV1,BergeS4,BergeSS3}).

In the two papers already mentioned exploring non-equilibrium scalar dynamics, only systems contained in a fixed volume box were
considered.  The isotropization of the
pressure tensor was therefore not relevant. To address this issue, we shall extend here our
study of scalar $\phi^4$ field theory to a system that
expands longitudinally.  In this case, one expects naively that the
longitudinal expansion produces a red-shifting of the longitudinal
momenta thereby leading to a small longitudinal pressure that
decreases faster than the transverse one. If this conclusion was right, hydrodynamics would not be applicable to the
description of such an expanding system. The main question we want to
address with this extension is whether the instabilities can keep the
longitudinal pressure close to the transverse one, despite the
longitudinal expansion. A study of these instabilities for an
expanding system has been presented in ref.~\cite{BergeBS1}, but the
main focus of this work was not on isotropization.

We will mimic closely, in this scalar theory framework, the Color
Glass Condensate description of high energy heavy ion collisions by
assuming that the classical background field is independent of
rapidity. The fluctuations that are superposed on this classical field
at the initial time are the scalar analogues of the spectrum derived
for QCD in \cite{FukusGM1,DusliGV1}. The derivation of this spectrum
in the much simpler scalar case is done in section
\ref{sec:specfluc}. In the section \ref{sec:lattice}, we discuss the
lattice discretization of the model. The time evolution of the
particle distribution is studied in the section
\ref{sec:occupnumber}. This discussion also shows why it is important
to have a very small lattice spacing in the rapidity direction.  The
time evolution of the transverse and longitudinal pressures is studied
in the section \ref{sec:enertensor}.  We conclude that the
instabilities are indeed able to make the longitudinal pressure nearly
equal to the transverse one.  In the section \ref{sec:hydro}, we
compare the results obtained here in classical statistical field
theory with expectations from viscous hydrodynamics. Section
\ref{sec:conclusion} is devoted to conclusions and to an outlook into
the application of this approach to QCD. Some technical notes and
complements are relegated to three appendices.

\section{Spectrum of fluctuations in an expanding
  system} \label{sec:specfluc} Following refs.~\cite{DusliEGV1,EpelbG1},
we consider a massless real scalar field theory with a quartic
coupling, whose Lagrangian density is given by
\begin{equation}
\mathcal{L}(\phi)=\frac{1}{2}\left(\partial^{\mu}\phi\right)^2
-\underbrace{\frac{g^2}{4!}\phi^4}_{V(\phi)}\;.
\label{eq:lag}
\end{equation}
The kinematics of a high energy nuclear collision is mirrored by picking a preferred spatial direction -- the $z$ direction in  
this paper -- to be the collision axis. For collisions in the high energy limit, the problem is
invariant under finite Lorentz boosts in this direction.  A natural system of coordinates in which this invariance takes a
simple form are the proper time $\tau$ and the rapidity $\eta$ coordinates that are
related to the usual Cartesian coordinates by the transformations
\begin{equation}
  \tau=\sqrt{t^2-z^2}\;,\quad 
  \eta=\frac{1}{2}\ln\left(\frac{t+z}{t-z}\right)
  \; ,
\end{equation}
with the transverse coordinates $x$ and $y$ unchanged.  In this
coordinate system, the d'Alembertian reads
\begin{equation}
\square
={\partial^2_\tau}+\frac{1}{\tau}{\partial_\tau}
-\frac{1}{\tau^2}{\partial^2_\eta}-{\bs\nabla}_{\perp}^2\;.
\end{equation}
Longitudinal boosts preserve $\tau$ and $\x_\perp$, and shift the
rapidity by a constant. Thus, the invariance under longitudinal boosts
implies that physical quantities are independent of the rapidity
$\eta$.

\subsection{Classical background field}
In the CGC description of heavy ion collisions, the leading order
approximation is given by the retarded classical solution of the
Yang-Mills equations. In subsequent orders, this classical
field plays the role of a background field whereby all the
propagators and vertices are dressed. This classical background  field
is boost invariant. Thus, in our scalar analog, we will require 
the background field to be independent of $\eta$.  The
classical equation of motion for such a field $\varphi$ is
\begin{equation}
\ddot\varphi+\frac{1}{\tau}{\dot\varphi}
-{\bs\nabla}_\perp^2\varphi+V'(\varphi)
=0\; ,
\end{equation}
where the dot denotes a derivative with respect to the proper time
$\tau$. Note that we only consider here the evolution of the system
after the collision (for $\tau>0$). Because the sources that 
describe the colliding projectiles have support only on the axis
$z=\pm t$,  the field evolution is not driven by these sources; instead, they provide initial
conditions for the field at $\tau=0^+$.

To ascertain the nature of the initial conditions at $\tau=0^+$, let us neglect for the time being the interaction term
$V'(\varphi)$ and decompose the $\x_\perp$ dependence of the field in
Fourier modes as
\begin{equation}
\varphi(\tau,\x_\perp)\equiv \int\frac{\d^2\k_\perp}{(2\pi)^2}\;
\varphi_{\k_\perp}(\tau)\; e^{i\k_\perp\cdot\x_\perp}\; .
\end{equation}
The Fourier coefficients obey the equation
\begin{equation}
\ddot\varphi_{\k_\perp}+\frac{1}{\tau}\dot\varphi_{\k_\perp}
+k_\perp^2\varphi_{\k_\perp}
=0\; .
\end{equation}
This is a Bessel equation of index $n=0$ and its general solution is
of the form
\begin{equation}
\varphi_{\k_\perp}(\tau)
=
a_{\k_\perp}\,J_0(k_\perp\tau)+b_{\k_\perp}\,Y_0(k_\perp\tau)\; ,
\end{equation}
where $a_{\k_\perp}$ and $b_{\k_\perp}$ are constants. Only the
term that contains $J_0$ can smoothly approach the limit $\tau\to
0^+$. (If we keep the $Y_0$ term, the field would diverge as
$\ln(1/\tau)$ and its derivative as $1/\tau$ in this limit.)

Therefore the solutions that are regular near the origin are of the form,
\begin{equation}
\varphi(\tau,\x_\perp)=\int\frac{\d^2\k_\perp}{(2\pi)^2}\;
e^{i\k_\perp\cdot\x_\perp}\; a_{\k_\perp}\;J_0(k_\perp\tau)\; ,
\end{equation}
where the $a_{\k_\perp}$ coefficients are just the Fourier transform
of the field at $\tau=0$
\begin{equation}
\varphi(0^+,\x_\perp)=\int\frac{\d^2\k_\perp}{(2\pi)^2}\;
e^{i\k_\perp\cdot\x_\perp}\; a_{\k_\perp}\; .
\end{equation}
Note that its derivative tends to zero at the origin,
\begin{equation}
\lim_{\tau\to 0^+}\dot\varphi(\tau,\x_\perp)=0\; .
\end{equation}

The interaction term $V'(\varphi)$ that we neglected in this
discussion introduces some mixing among the Fourier modes and
modifies the evolution of the solution. However this modification is
negligible at short times because the evolution of the field at very
short times is dominated by the {\sl expansion rate}  which is
much larger than the {\sl interaction rate}. Therefore the solution
that we find by neglecting the interaction term remains valid
in the immediate vicinity of the origin\footnote{How far in $\tau$ this
approximation is valid depends on the magnitude of the field and of
the coupling constant $g$, both of which determine the strength of the neglected non-linear terms.}.

\subsection{Spectrum of fluctuations}
We want now to perform for this scalar model the resummation described
in ref.~\cite{FukusGM1,DusliGV1} for the QCD case. This resummation amounts to
allowing the initial condition for the classical field to
fluctuate and to average observables over these
fluctuations. Following \cite{DusliGV1}, we can construct the Gaussian
ensemble of initial conditions by superposing on the background field
$\varphi$ (independent of $\eta$)  a fluctuating $\eta$ dependent term, with the result expressed as 
\begin{equation}
\phi=\varphi+\int \d\mu_{_K}\;\Big[c_{_K}\,a_{_K}+c_{_K}^*\,a_{_K}^*\Big]\; .
\label{eq:fluct}
\end{equation}
Here the $a_{_K}$ are mode functions propagating on top of the
classical background field $\varphi$ that obey the equation
of motion
\begin{equation}
\ddot{a}_{_K}+\frac{1}{\tau}\dot{a}_{_K}-\frac{1}{\tau^2}\partial_\eta^2 a_{_K}
-{\bs\nabla}_\perp^2 a_{_K}+V''(\varphi) a_{_K}=0\; .
\label{eq:flucteom}
\end{equation}
This is a linear equation and the set of its solutions is a vector
space. The symbol $K$ denotes collectively all the labels that are
necessary in order to index the basis for this vector space and
$\d\mu_{_K}$ is the measure necessary to integrate over this space. In eq.~(\ref{eq:fluct}) the $a_{_K}$'s must be positive
frequency solutions and they must form an orthonormal basis with
respect to the inner product
\begin{equation}
\big(a\big|b\big)\equiv i \tau\int \d\eta\, \d^2\x_\perp\;(a^*\dot{b}-\dot{a}^*b) 
\; .\label{eq:scalarprod}
\end{equation}
This means that one must have
\begin{equation}
\big(a_{_K}\big|a_{_L}\big)=\delta_{_{KL}}\; ,
\end{equation}
where the $\delta$-function on the space of the indices $K$ is normalized to ensure
\begin{equation}
\int \d\mu_{_K}\;\delta_{_{KL}}=1\; .
\end{equation}
The coefficients $c_{_K}$ are Gaussian random complex numbers whose
variance is defined by the equations,
\begin{eqnarray}
\big<c_{_K}c_{_L}\big>&=&\big<c_{_K}^*c_{_L}^*\big>=0
\nonumber\\
\big<c_{_K}c_{_L}^*\big>&=&\frac{1}{2}\delta_{_{KL}}\; .
\end{eqnarray}
We emphasize that the normalization condition of the fluctuations
$a_{_K}$ depends on how we define the integration measure so that at
the end of the day the ensemble of $\phi$'s defined by
eq.~(\ref{eq:fluct}) does not depend on this arbitrary choice.

\subsubsection{Mode functions}
Let us now solve eq.~(\ref{eq:flucteom}) which determines the mode
functions. We wish only to obtain its solutions at small times $\tau$, therefore we shall neglect the time dependence\footnote{Since we saw in the previous section that its
time derivative should vanish when $\tau\to 0^+$.} of the background
field $\varphi$ and replace it by its limit at $\tau\to0^+$,
\begin{equation}
\varphi_0(\x_\perp)\equiv\lim_{\tau\to0^+}\varphi(\tau,\x_\perp)\; .
\end{equation}
Since $\varphi_0$ does not depend on the rapidity $\eta$, we can
look for solutions that have a well defined wave-number in the variable
$\eta$ of the form
\begin{equation}
a_\nu(\tau,\eta,\x_\perp)
=
e^{i\nu\eta}\; b_\nu(\tau,\x_\perp)\; .
\end{equation}
The function $b_\nu$ now obeys
\begin{equation}
\ddot{b}_\nu+\frac{1}{\tau}\dot{b}_\nu
+\frac{\nu^2}{\tau^2}b_\nu
-{\bs\nabla}_\perp^2 b_\nu+V''(\varphi_0) b_\nu=0\; .
\label{eq:flucteom1}
\end{equation}
We note that the $\x_\perp$ dependence of the solution is controlled by
the operator $-{\bs\nabla}_\perp^2+V''(\varphi_0)$. This operator is
real and symmetric and can therefore be diagonalized over a basis of
orthogonal functions $\chi_\k(\x_\perp)$,
\begin{eqnarray}
\left[-{\bs\nabla}_\perp^2+V''(\varphi_0)\right]\,\chi_\k(\x_\perp)
&=&
\lambda_\k^2\,\chi_\k(\x_\perp)
\nonumber\\
\int \d^2\x_\perp\;\chi_\k^*(\x_\perp)\chi_\l(\x_\perp)&=&\delta_{\k\l}\; .
\label{eq:eigenvalue}
\end{eqnarray}
The $\delta$-function $\delta_{\k\l}$ is defined so that its integral
over the measure $\d\mu_{\k}$ in $\k$-space is 
\begin{equation}
\int \d\mu_\k\;\delta_{\k\l}=1\;.
\end{equation}

The next step is to decompose the $\x_\perp$ dependence of $b_\nu$ on
the basis provided by the $\chi_\k(\x_\perp)$'s. This amounts to
looking for fluctuations that are of the form
\begin{equation}
a_{\nu\k}(\tau,\eta,\x_\perp)
=
e^{i\nu\eta}\;
\chi_\k(\x_\perp)\;
b_{\nu\k}(\tau)\; .
\end{equation}
The remaining function $b_{\nu\k}(\tau)$ depends only on time and must
obey the following equation
\begin{equation}
\ddot{b}_{\nu\k}+\frac{1}{\tau}\dot{b}_{\nu\k}
+\frac{\nu^2}{\tau^2}b_{\nu\k}
+\lambda_\k^2 b_{\nu\k}=0\; .
\label{eq:flucteom2}
\end{equation}
This is a Bessel equation whose general solution is a linear
combination of the Hankel functions\footnote{One could also use the
  Bessel functions $J_{i\nu}(\lambda_\k\tau)$ and
  $Y_{i\nu}(\lambda_\k\tau)$ as the two independent solutions, but
  they are less convenient in the problem at hand because they are
  both mixtures of positive and negative frequency components.}
$H_{i\nu}^{(1)}(\lambda_\k\tau)$ and
$H_{i\nu}^{(2)}(\lambda_\k\tau)$. The latter is the positive frequency
solution, as can be seen from its asymptotic behavior at large
time\footnote{For large arguments, one has\begin{equation*}
H_{i\nu}^{(2)}(\tau)\empile{\approx}\over{\tau\to+\infty}
\sqrt{\frac{2}{\pi\tau}}\;e^{-i(\tau-i\pi\nu/2-\pi/4)}\; .
\end{equation*}}. Thus the $a_{_K}$'s in eq.~(\ref{eq:fluct}) can be
chosen as
\begin{equation}
a_{\nu\k}(\tau,\eta,\x_\perp)
=
\alpha_{\nu\k}\;e^{i\nu\eta}\;
\chi_\k(\x_\perp)\;H_{i\nu}^{(2)}(\lambda_\k\tau)\; .
\end{equation}

We need only to determine the constant prefactor $\alpha_{\nu\k}$ in order to ensure that these solutions are properly normalized. (They are, by construction, 
already mutually orthogonal.) To obtain this, we can compute
the inner product
\begin{eqnarray}
\big(a_{\nu\k}\big|a_{\mu\l}\big)&=&
 i \tau\int \d\eta\, \d^2\x_\perp\;(a^*_{\nu\k}\dot{b}_{\mu\l}-\dot{a}^*_{\nu\k}b_{\mu\l}) 
 \nonumber\\
&=&
i\tau\,2\pi\delta(\nu-\mu)\delta_{\k\l}\,\left|\alpha_{\nu\k}\right|^2\,
\underbrace{H_{i\nu}^{(2)*}(\lambda_\k\tau)\stackrel{\leftrightarrow}{\partial}_\tau
H_{i\nu}^{(2)}(\lambda_\k\tau)}_{-4i e^{-\pi\nu}/\pi\tau}
\nonumber\\
&=&8\delta(\nu-\mu)\delta_{\k\l}\,\left|\alpha_{\nu\k}\right|^2\,e^{-\pi\nu}\; ,
\end{eqnarray}
where the  identity used to go from the second to the third line is the Wronskian between $H_{-i\nu}^{(1)}=\big(H_{i\nu}^{(2)}\big)^*$
and $H_{i\nu}^{(2)}$.  If we choose the integration measure over the
indices $\nu,\k$ to be $\d\mu_{_K}\equiv\frac{\d\nu}{2\pi}\,\d\mu_\k$,
then this inner product should be $2\pi\delta(\nu-\mu)\delta_{\k\l}$,
which can be satisfied by taking
\begin{equation}
\alpha_{\nu\k}=\frac{\sqrt{\pi}e^{\pi\nu/2}}{2}\; .
\end{equation}

\subsubsection{Final form of the spectrum}
Let us now summarize the results of this section. We showed that the fluctuating fields
defined in eq.~(\ref{eq:fluct}) should have the explicit form
\begin{equation}
\phi(\tau,\eta,\x_\perp)
=\varphi(\tau,\x_\perp)
+
\frac{\sqrt{\pi}}{2}\int \frac{\d\nu}{2\pi}\,\d\mu_\k\;e^{\pi\nu/2}
c_{\nu\k}\;
e^{i\nu\eta}\;
\chi_\k(\x_\perp)\;H_{i\nu}^{(2)}(\lambda_\k\tau)
+\mbox{c.c.}\; ,
\label{eq:fluct1}
\end{equation}
where the random coefficients in this linear superposition have the variance
\begin{equation}
\big<c_{\nu\k}c_{\mu\l}\big>=0\; ,\quad\big<c_{\nu\k}c_{\mu\l}^*\big>=\pi\delta(\nu-\mu)\delta_{\k\l}\; .
\end{equation}
Thus far, we haven't been very explicit about the nature of the index $\k$
that labels the eigenfunctions of the operator
$-{\bs\nabla}_\perp^2+V''(\varphi_0)$. This label is a 2-dimensional
continuous index that plays the same role as the transverse momentum
in the free case\footnote{In the free case, the corresponding measure
  and delta function would thus be defined as
  \begin{equation*}
    \d\mu_\k\equiv \frac{\d^2\k_\perp}{(2\pi)^2}\;,\quad \delta_{\k\l}\equiv (2\pi)^2\delta(\k_\perp-\l_\perp)\; .
  \end{equation*}
}.

The fields defined by eq.~(\ref{eq:fluct1}) do not seem to have a well
defined limit when $\tau\to 0^+$, because the Hankel functions
oscillate like $\tau^{\pm i\nu}$ as $\tau$ approaches 0 and their
derivatives diverge. This forces us to start the time evolution
at a small but finite time $\tau_0$, rather than at exactly
$\tau=0^+$. In this respect the fields with rapidity dependent
fluctuations superposed on the background field are qualitatively different from the background field, for
which starting from $\tau=0^+$ poses no difficulty. However, the finite $\tau_0$ that we are
forced to introduce as a consequence, fortuitously has no effect on the physical results. Indeed
since eq.~(\ref{eq:fluct1}) is a solution of the equations of motion
at small $\tau$, one can prove that physical quantities computed at
some later time by averaging over these $\phi$'s will be independent
of the choice of $\tau_0$ provided it is chosen to be small enough. This
will be checked in the section \ref{sec:initime}.

\section{Lattice implementation}
\label{sec:lattice}

An explicit determination of the fields in eq.~(\ref{eq:fluct1}) requires that we solve the eigenvalue equation in eq.~(\ref{eq:eigenvalue}). Because this expression depends nonlinearly on the initial background field, the solution has to be determined numerically. In this section, we will outline the numerical procedure followed in the solution.

\subsection{Discretization}
\begin{figure}[htbp]
\begin{center}
\resizebox*{!}{6cm}{\includegraphics{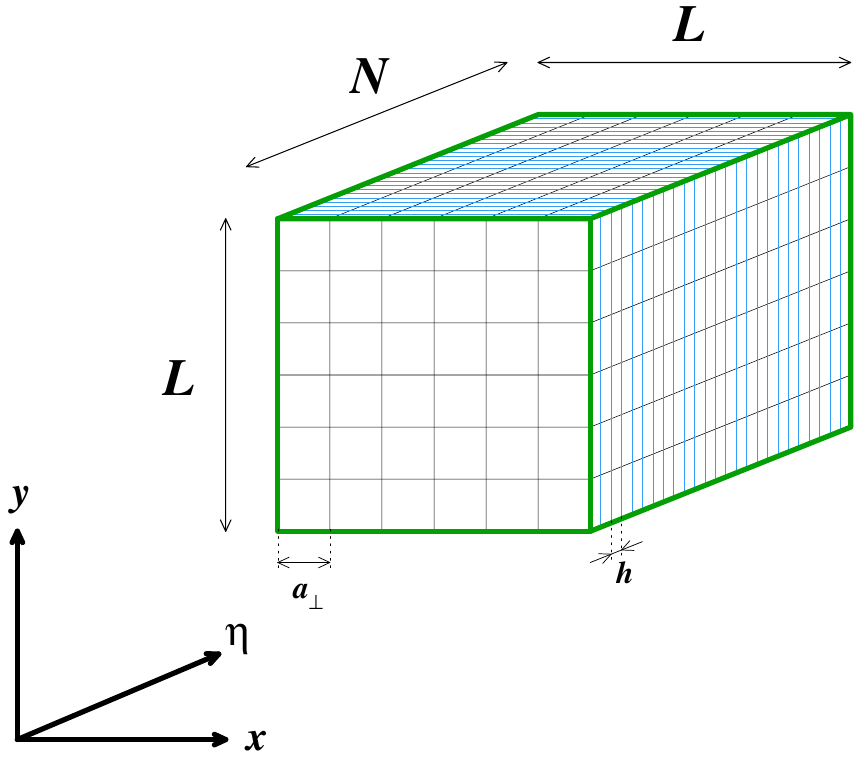}}
\end{center}
\caption{\label{fig:lattice}Details of the lattice discretization used
  in our numerical implementation.}
\end{figure}
In view of a numerical implementation, we discretize the problem
as follows:
\begin{itemize}
\item[{\bf i.}] The transverse plane $\x_\perp$ becomes a $L\times L$
  lattice, and the lattice sites are labeled by a pair of integers
  $(i,j)$ (that range from 0 to $L-1$). We use periodic boundary
  conditions.  The indices $i,j$ are thus defined modulo $L$. We follow the 
  convention to set the transverse lattice spacing to unity,
  $a_\perp=1$. All the other dimensionful
  quantities of the problem are then expressed in units of the appropriate
  power of the transverse lattice spacing. For example, the values of the time
  quoted later on in the paper should be understood as values in units of $\tau/a_\perp$.

\item[{\bf ii.}] We consider a unit slice of rapidity, discretized in
  $N$ intervals, labeled by an integer $n$ that ranges from 0 to
  $N-1$. We also adopt periodic boundary conditions here. This is justified for high energy collisions because we expect
  observables to be invariant by translation in $\eta$ for the $\eta$ range of interest for thermalization. The index $n$
  is thus defined modulo $N$. The lattice spacing in the rapidity
  direction is dimensionless and its value is $h\equiv 1/N$.

\item[{\bf iii.}] We discretize the second derivatives in $\eta$ and
  $\x_\perp$ by symmetric finite differences as 
  \begin{eqnarray}
    \partial_\eta^2\phi &\to& \frac{1}{h^2}\left(\phi_{n+1ij}+\phi_{n-1ij}-2\phi_{nij}\right)
    \nonumber\\
    {\bs\nabla}_\perp^2\phi &\to& \phi_{ni+1j}+\phi_{ni-1j}
    +\phi_{nij+1}+\phi_{nij-1}
    -4\phi_{nij}\; .
  \end{eqnarray}

\item[{\bf iv.}] Time remains a continuous
  variable\footnote{Time steps in the numerical implementation can be made as small (and even changed dynamically) as
    needed to ensure the desired accuracy.}.
\end{itemize}

On the lattice, the eigenvalue problem for finding the $\chi_\k$'s is
transformed into the problem of finding the eigenvalues and
eigenvectors of a real symmetric matrix, since
\begin{equation}
\left[-{\bs\nabla}_\perp^2+V''(\varphi_0)\right]\chi \;\to\;(D\chi)_{ij}\equiv
4\chi_{ij}-\chi_{i+1j}-\chi_{i-1j}-\chi_{ij+1}-\chi_{ij-1}+V''_{ij}\chi_{ij}\; .
\end{equation}
Thus, the $L^2\times L^2$ matrix $D_{ij,kl}$ that we must diagonalize
has the following expression
\begin{equation}
D_{ij,kl} = (4+V''_{ij})\delta_{ik}\delta_{jl}
-(\delta_{i+1k}+\delta_{i-1k})\delta_{jl}
-\delta_{ik}(\delta_{j+1l}+\delta_{j-1l})\; .
\end{equation}
This diagonalization will provide us a set of $L^2$ eigenvectors
$\chi^{(p)}_{ij}$ such that
\begin{eqnarray}
\sum_{kl}D_{ij,kl}\,\chi^{(p)}_{kl} &=& \lambda_{(p)}^2\chi^{(p)}_{ij}
\nonumber\\
\sum_{ij}\chi^{(p)*}_{ij}\chi^{(q)}_{ij} &=& L^2 \delta_{pq}\; .
\end{eqnarray}

Once this is done, the fields of eq.~(\ref{eq:fluct1}) can be written as
\begin{equation}
\phi_{nij}(\tau)
=
\varphi_{ij}(\tau)
+
\frac{\sqrt{\pi}}{2NL^2 h}\sum_{v=0}^{N-1}e^{\pi\nu/2}
\sum_{p=0}^{L^2-1}
\Big[ c_{vp}\;
e^{\frac{2i\pi vn}{N}}\;
\chi^{(p)}_{ij}
\;H_{i\nu}^{(2)}(\lambda_{(p)}\tau)+\mbox{c.c.}\Big]\; ,\label{eq:discspect}
\end{equation}
where $\nu$ is an eigenvalue of the discretized operator 
$\partial_\eta^2$
\begin{equation}
\nu^2 = \frac{2}{h^2}\left[1-\cos\left(\frac{2\pi v}{N}\right)\right]\; ,
\end{equation}
and where the $c_{vp}$ are random Gaussian numbers with the variance
\begin{equation}
\big<c_{np}c_{mq}^*\big>=\frac{NL^2 h}{2}\delta_{nm}\delta_{pq}\; .
\end{equation}

\subsection{Special case of a uniform background field}
If the background field $\varphi(\tau,\x_\perp)$ is independent of
$\x_\perp$, the diagonalization of the matrix $D_{ij,kl}$ is
trivial because $V''_{ij}$ is just a constant mass term
$m^2=g^2\frac{\varphi(0^+)^2}{2}$, independent of position in the
transverse plane. The eigenfunctions are thus plane waves labeled by two
integers $p,q$ between 0 and $L-1$,
\begin{equation}
\chi^{(pq)}_{ij}=e^{\frac{2i\pi(pi+qj)}{L}}\; ,
\end{equation}
and the corresponding eigenvalues are
\begin{equation}
\lambda^2_{(pq)}
=
m^2+2\left[2-\cos\left(\frac{2\pi p}{L}\right)-\cos\left(\frac{2\pi q}{L}\right)\right]\; .
\end{equation}

\subsection{Independence with respect to the initial
  time} \label{sec:initime} As noted previously, one cannot take the
time to zero in equation (\ref{eq:fluct1}) due to the oscillatory
behavior of the Hankel functions. Thus one must choose a small initial
time $\tau_0$, compute the initial fields at this time, and proceed
with the time evolution from there. However, since this initial time
is arbitrary, observables should not depend on it. One could view this
question within the framework of the renormalization group:
observables can be made independent of $\tau_0$ by making the ensemble
of initial fields depend on $\tau_0$ in such a way as to compensate
for the fact that the time evolution starts at $\tau_0$. In fact, the
time dependence of eq.~(\ref{eq:fluct1}) is precisely what is needed
to achieve this.

\begin{figure}[htbp]
\begin{center}
\resizebox*{10cm}{!}{{\includegraphics{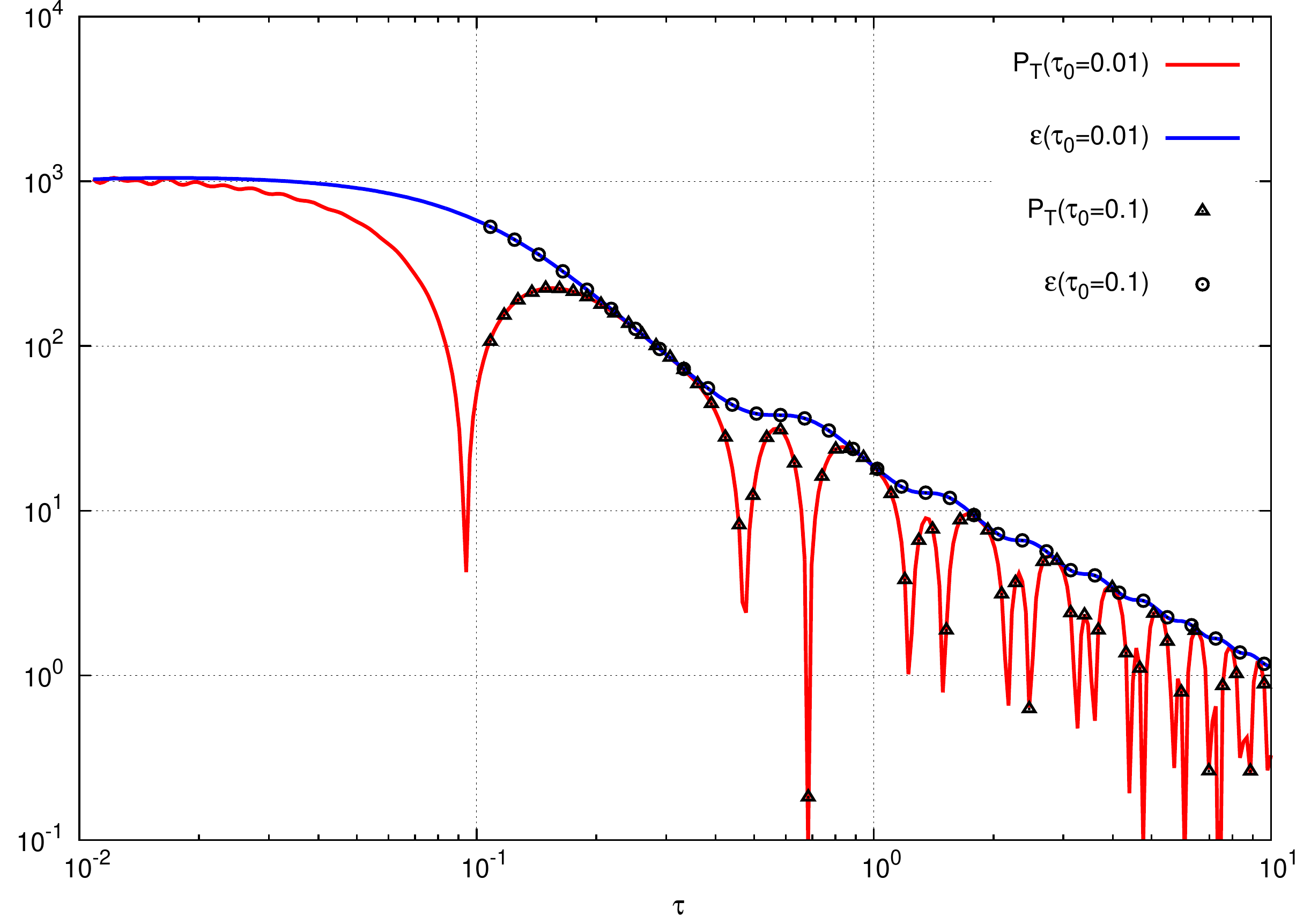}}}
\end{center}
\caption{\label{fig:checktau}Time evolution of some components of the
  energy-momentum tensor, for two different initial times $\tau_0=0.01$
  (solid lines) and $0.1$ (points). This computation was done on a
  $20\times20\times20$ lattice with $256$ field configurations. All
  the plots presented in this paper (with the exception of fig.~\ref{fig:g2dep} in appendix B) correspond to coupling
  constant $g=4$.}
\end{figure}

In figure \ref{fig:checktau}, we show the time evolution of the energy
density $\varepsilon$ and transverse pressure $P_T$ (see
eqs.~(\ref{eq:pressures}) for explicit expressions in terms of the
scalar field) of the system for two different initial times,
$\tau_0=10^{-2}$ and $\tau_0^\prime=10^{-1}$.
One can see that, for the times that are common to the two evolutions,
these physical quantities do not depend on the time at which the
system is initialized. One should add a word of caution here.  This
statement is true only as long as the initial time $\tau_0$ is small
compared to the physical time scales in the problem (for example, the period
of the oscillations of the pressure in the fig.~\ref{fig:checktau}).
Note also that the classical background $\varphi$ itself in
eq.~(\ref{eq:fluct1}) evolves from $\tau_0$ to $\tau_0^\prime$,
although this is a small effect if both initial times are small.

\subsection{Unstable modes}
A crucial aspect of dynamics in this theory is the presence of
unstable modes that are amplified by parametric resonance (see also
ref.~\cite{BergeBS1} for a discussion of this question). We saw in
\cite{DusliEGV1} that the resonance band for a constant volume is
fixed and that its boundaries are proportional to the amplitude of the
background classical field. The situation becomes more complicated
when the system is free to expand in the $z$ direction.

For simplicity, let us assume
that the background field $\varphi$ is spatially
homogeneous\footnote{This choice is the one for which the structure of
  the resonance band is the simplest. See for instance
  \cite{GreenKLS1}.}. The linearized equation of motion for a small
fluctuation of wave numbers $\nu$ and $\k_\perp$ propagating over this
background reads
\begin{equation}
\ddot{a}+\frac{1}{\tau}\,\dot{a}
+\left(k_\perp^2+\frac{\nu^2}{\tau^2}+\frac{g^2}{2}\varphi^2(\tau)\right)\,a=0\; .
\end{equation}
When the background field is not present ($\varphi\equiv 0$), the
solutions of this equation are the Bessel functions $J_{i\nu}(k_\perp
\tau)$ and $Y_{i\nu}(k_\perp \tau)$. At late times, they oscillate
with an amplitude that decreases like $\tau^{-1/2}$. When
$\varphi\not=0$, one can check (numerically) at any given time whether
the fluctuation of wavenumbers $\nu,\k_\perp$ has the expected
magnitude, namely, if they were decreasing at the pace of the above Bessel
functions, or if they are larger than expected. If the latter is the case, one interprets this mode to have been amplified by
the resonance in the course of its evolution. 

\begin{figure}[htbp]
\begin{center}
\resizebox*{10cm}{!}{{\includegraphics{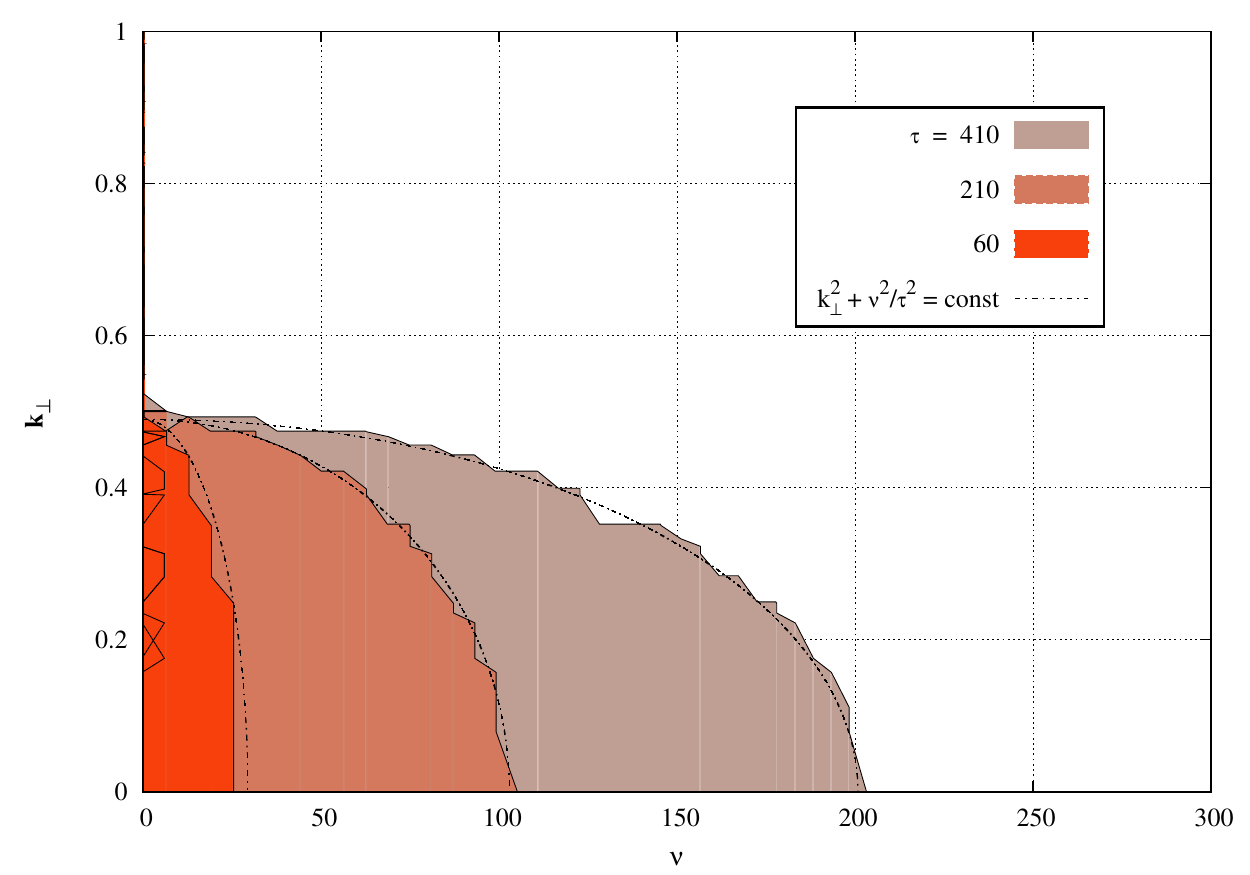}}}
\end{center}
\caption{\label{fig:reso}Modes that have been amplified by the
  instability at various times in the evolution.}
\end{figure}

In  fig. \ref{fig:reso}, we plot modes that are unexpectedly large at the time
$\tau$ for three values of $\tau$.
The number of modes that have been affected by
the resonance increases with time. Moreover, the boundary of this
region is well reproduced by a line of constant
$k_\perp^2+\nu^2/\tau^2$.  This fact is easy to understand
semi-quantitatively from what we know about the resonance in the
fixed volume case. There the resonant modes form a narrow band
\begin{equation}
R_-<k_\perp^2+k_z^2 < R_+\; ,
\end{equation}
where $R_\pm$ are two numbers proportional to the amplitude squared of
the background field. The resonance occurs due to a special tuning between
the ``effective mass''  $k^2$ of a mode and its coupling to
the oscillating background field. In the expanding case, it is the
combination $k_\perp^2+\nu^2/\tau^2$ that plays the role of a mass
term for the mode. For large times such that the variations of
$1/\tau^2$ are slow compared to the oscillations of the background
field we can justly transpose to this case the resonance condition obtained in the
non-expanding case. We then get the resonance condition,
\begin{equation}
R_-<k_\perp^2+\frac{\nu^2}{\tau^2} < R_+\; .
\end{equation}
These inequalities define a domain that has the shape of a narrow
corona in the $\nu,k_\perp$ plane. The main effect of the expansion is
that this corona expands with time, starting very close to the $\nu=0$
axis and progressively moving towards larger values of $\nu$.
Therefore the modes that are unstable are not the same at all
times. Moreover, a given $\nu,\k_\perp$ mode can only be resonant
during a finite amount of time while the resonance band is sweeping
through it. During this time, it is amplified by some possibly large
but always finite factor.

From this discussion it is easy to understand fig. \ref{fig:reso}. The shaded areas represent all the modes that have
been resonant at some point between the initial time and the time
$\tau$ of interest. This domain is the addition of the locations of
the resonance band for all the times up to $\tau$. In this
interpretation, the boundary of the domain (of equation
$k_\perp^2+\nu^2/\tau^2=\mbox{const}$) corresponds to the modes that
are just starting to become resonant at the time $\tau$. Note that the
modes that become resonant at late times are not amplified as strongly
as those that resonate early: indeed, as known already for a fixed box, the Lyapunov exponent is
proportional to the amplitude of the background field--which in this case decreases with time.

\section{Occupation number} \label{sec:occupnumber}

In the next two sections, we will discuss results from our lattice
simulations. In this section, we will discuss results for the
occupation number. In the following section, results for components
of the stress-energy tensor will be presented.
\subsection{Definition}
We begin with the
Fourier decomposition of a free scalar field operator\footnote{The seemingly peculiar normalization factors in this formula will
become clear after eq.~(\ref{eq:a-norm}).}  in the system of
coordinates $(\tau,\eta,\x_\perp)$
\begin{equation}
\widehat\phi(\tau,\eta,\x_{\perp})
=
\frac{\sqrt{\pi}}{2}\int \frac{\d^2\k_\perp}{(2\pi)^2}\frac{\d\nu}{2\pi}
\;
e^{{\pi\nu}/{2}}\,
H_{i\nu}^{(2)}(k_\perp\tau)\;
\widehat{a}_{\nu\k_\perp}\;e^{i\nu\eta}\,e^{i\k_{\perp}\cdot\x_{\perp}}+\mbox{h.c.}\;.
\end{equation}
One may check immediately that this expression is a solution of the free field equation of
motion. This equation can be inverted to obtain creation and
annihilation operators expressed in terms of the field as
\begin{equation}
\widehat{a}_{\nu\k_\perp} 
= 
i\tau e^{{\pi\nu}/{2}}\frac{\sqrt{\pi}}{2}
\int \d^2\x_{\perp}\d\eta\; 
e^{-i\k_{\perp}\cdot\x_{\perp}}\,e^{-i\nu\eta}
H_{i\nu}^{(2)*}(k_\perp\tau)
\stackrel{\leftrightarrow}{\partial_{\tau}}\widehat\phi(\tau,\eta,\x_{\perp})\; .
\label{eq:inv-form}
\end{equation}
The normalization in these formulas has been chosen to satisfy
\begin{equation}
\left[\widehat{a}_{\nu\k_\perp},\widehat{a}^\dagger_{\nu'\l_\perp}\right]
=(2\pi)^3\delta(\nu-\nu')\delta(\k_\perp-\l_\perp)\; ,
\label{eq:a-norm}
\end{equation}
given the canonical commutation relations for the field operator.
When we set $\nu=\nu'$ and $\k_\perp=\l_\perp$, the delta functions on 
the right hand side become $S_\perp L_\eta$, where $S_\perp$ is the
transverse area of the system and $L_\eta$ the length of the rapidity
interval under consideration.

As explained in ref.~\cite{EpelbG1} (sections 4.1 and 4.2), it is
straightforward to obtain the expectation value of the symmetric
combination $a^\dagger a + a a^\dagger$ in the resummation scheme we
are using here. For creation and annihilation operators normalized as
in eq.~(\ref{eq:a-norm}), it is natural to define the occupation
number as 
\begin{equation}
1+2f_{\nu\k_\perp} \equiv \frac{1}{S_\perp L_\eta}
\left< 
\widehat{a}_{\nu\k_\perp}^{\dagger}\widehat{a}_{\nu\k_\perp}
+
\widehat{a}_{\nu\k_\perp}\widehat{a}_{\nu\k_\perp}^{\dagger}
\right >\;.
\end{equation}
From eq.~(\ref{eq:inv-form}), the occupation number
computed in our resummation scheme can be expressed as
\begin{equation}
f_{\nu\k_\perp}(\tau)= -\frac{1}{2}
+
\frac{\pi\tau^2 e^{\pi\nu}}{4 S_\perp L_\eta}
\left<\Big|
\int  \d^2\x_\perp\d\eta \;e^{-i\nu\eta}e^{-i\k_\perp\cdot\x_\perp}\;
H_{i\nu}^{(2)*}(k_\perp\tau)\stackrel{\leftrightarrow}{\partial}_\tau 
\phi(\tau,\eta,\x_\perp)\Big|^2\right>\; .
\label{eq:fk}
\end{equation}
Here $\phi(\tau,\eta,\x_\perp)$ is a classical solution of the field
equation of motion and the brackets denote the average over the
Gaussian ensemble of initial conditions.

A crucial aspect of eq.~(\ref{eq:fk}) is the prefactor $\tau^2
e^{\pi\nu}$ in front of the integral that ensures that each mode is
weighted correctly at all times despite the dilution due to the
longitudinal expansion of the system.  One can also check that this
formula gives a vanishing occupation number in the case of pure vacuum
fluctuations (i.e. with $\phi$ replaced by eq.~(\ref{eq:fluct1}) with
$\varphi\equiv0$). Moreover, we have checked that, when evolved in
time with the interacting equation of motion, this vacuum ensemble of
fields evolves in such a way as to give a vanishing occupation number
over all the time range of interest\footnote{If this evolution is
  carried to extremely large times, we may expect that these vacuum
  fluctuations will eventually ``thermalize'' \cite{Moore2}.}.

\subsection{Time evolution}
\label{sec:fk-evol}
\begin{figure}[htbp]
\begin{center}
\resizebox*{7cm}{!}{{\includegraphics{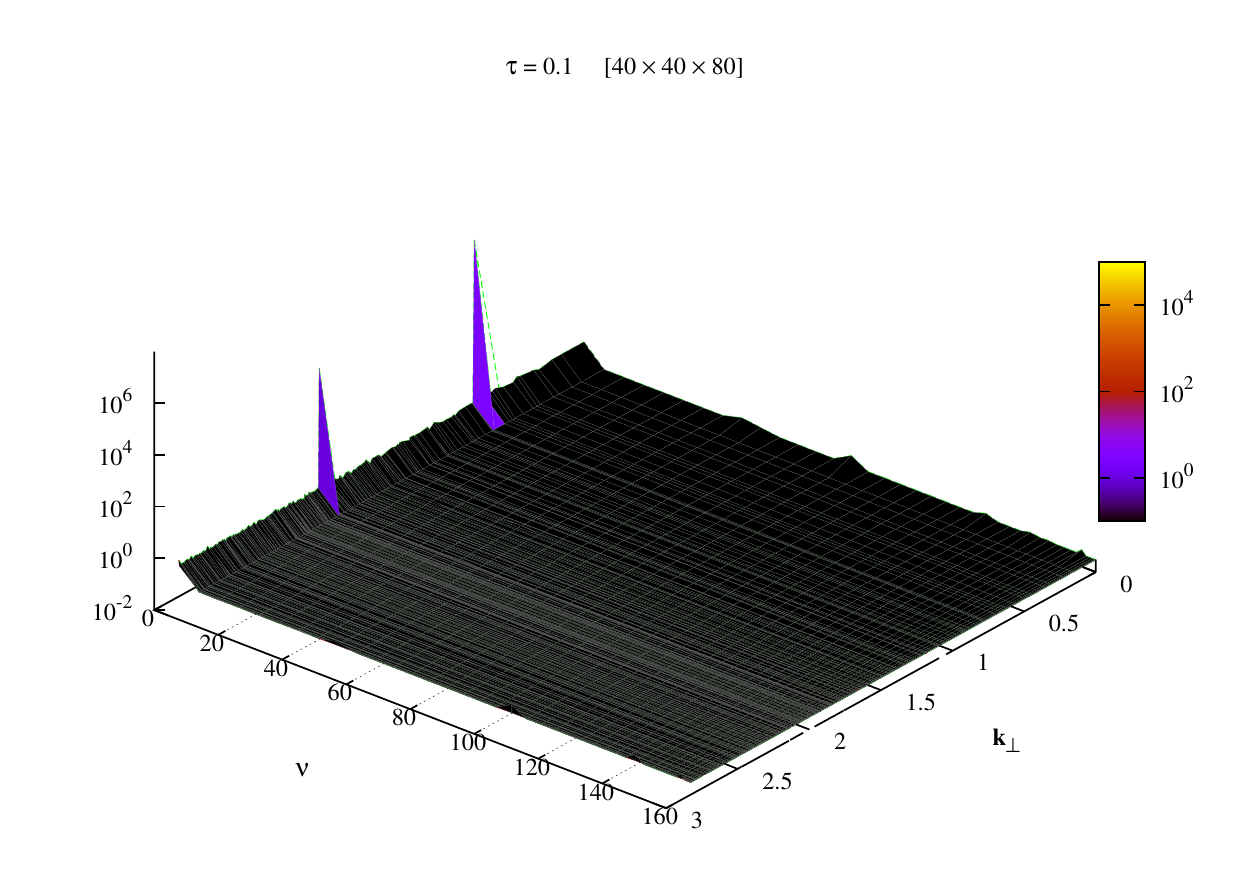}}}
\resizebox*{7cm}{!}{{\includegraphics{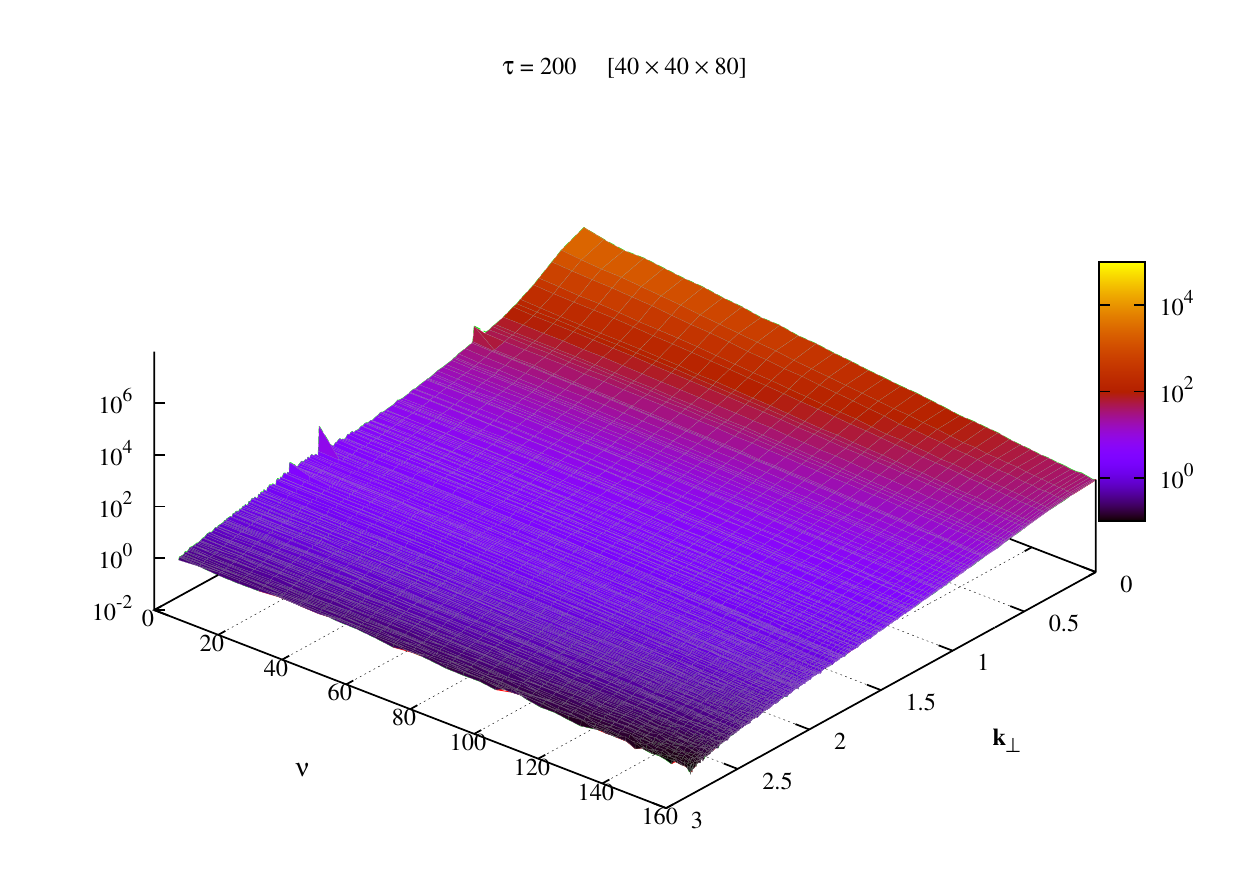}}}
\end{center}
\caption{\label{fig:fk80}Occupation number at times $\tau=10^{-1}$ and
  $\tau=200$, on a $40\times40\times80$ lattice.}
\end{figure}

In fig. \ref{fig:fk80}, we show the occupation number at two times,
computed on a $40\times40\times80$ lattice, for an initial classical
background field $\varphi_0(\x_\perp)$ at $\tau=0^+$. As an example,
we take a background field that initially contains only a single mode
$\k_\perp$,
\begin{equation}
\varphi_0(\x_\perp) = \varphi_0\times \cos(\k_\perp\cdot\x_\perp)\; .
\end{equation}
This is arguably a simplistic toy model of the classical background
field. In the Color Glass Condensate description of a heavy ion
collision, the classical background field would contain a range of
non-zero modes up to the saturation momentum. The coupling constant
used in this computation, and in all the plots shown in this paper,
is $g=4$. The value of the background field has been taken to
$\varphi_0=15$, and $\k_\perp=0.77$.
At the time $\tau=0.1$, a harmonic of the initial $\k_\perp$
mode has already started to grow in the distribution (see the second
peak in the left plot of the figure \ref{fig:fk80}), but its amplitude
is comparatively very small (note that the vertical axis is logarithmic). On
the other hand, the distribution in $\nu$ is still confined at $\nu=0$ because the fluctuating part of the fields is still very small
compared to the boost invariant classical background. The plot on the
right of the figure \ref{fig:fk80}, that represents the occupation
number at the time $\tau=200$, shows that a continuum of $\k_\perp$
and $\nu$ modes eventually get populated after some evolution.

This plot also illustrates the main difficulty in simulating an
expanding system.  Over time the distribution extends in the $\nu$
direction and eventually reaches the lattice cutoff in this
variable. The origin of this behavior can be easily understood. For
simplicity, let us assume that the system has reached local thermal
equilibrium and expands following ideal Bjorken
hydrodynamics. Then its energy density decreases like $\epsilon\sim
\tau^{-4/3}$ and its temperature decreases like
$T\sim\tau^{-1/3}$. The temperature is roughly the extent of the
support of the occupation number in the variables $k_\perp$ and
$k_z$. Therefore in these variables the support of the occupation
number tends to shrink slowly with time.  However at late times the
variables $k_z$ and $\nu$ are related by
\begin{equation}
k_z \sim \frac{\nu}{\tau}\; ,
\end{equation}
and therefore a shrinkage as $\tau^{-1/3}$ of the momentum $k_z$
corresponds to an increase with $\tau^{2/3}$ of the variable
$\nu$. Therefore if the system thermalizes\footnote{In contrast, if
  the system is free-streaming, $k_\perp\sim\mbox{const}$ and
  $k_z\sim \tau^{-1}$. This corresponds to a constant support both in
  $k_\perp$ and in $\nu$.} or approaches thermal
equilibrium, there will always be a time where the support of
$f_{\nu\k_\perp}$ hits the lattice cutoff in the variable $\nu$ no
matter how high this cutoff is.

This is a serious issue for our simulation because physical quantities
computed after this time will be contaminated by lattice
artifacts. This is the case in particular for components of the
pressure tensor as they are sensitive to the hardest populated modes
in the system. When the occupation number reaches the upper $\nu$
limit, the distribution in $k_z$ is artificially cut-off, which leads
to a reduction of the longitudinal pressure. If we were to pursue the
time evolution much beyond this time, the system would effectively
become 2-dimensional and we would get the incorrect result $P_{_L}\ll
P_\perp$.

\begin{figure}[htbp]
\begin{center}
\resizebox*{7cm}{!}{{\includegraphics{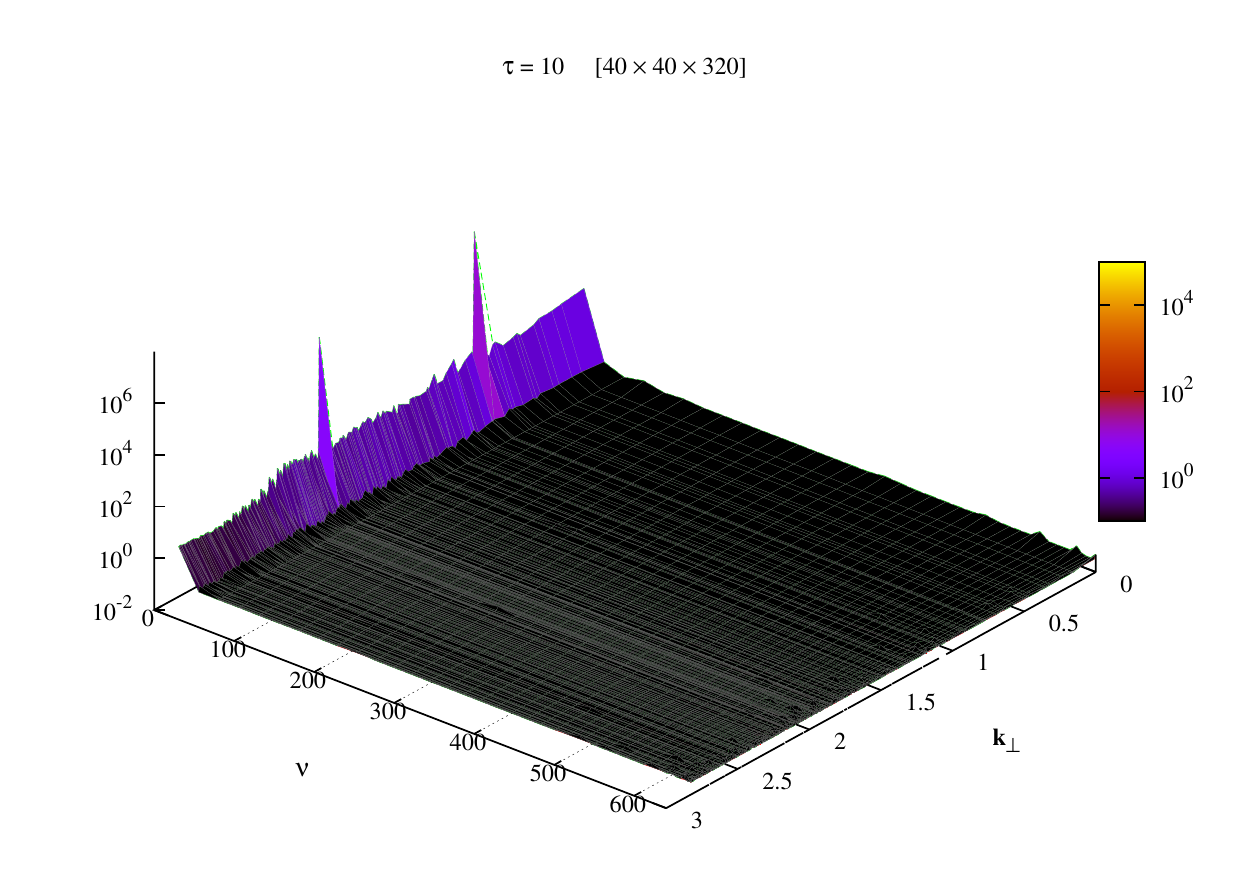}}}
\resizebox*{7cm}{!}{{\includegraphics{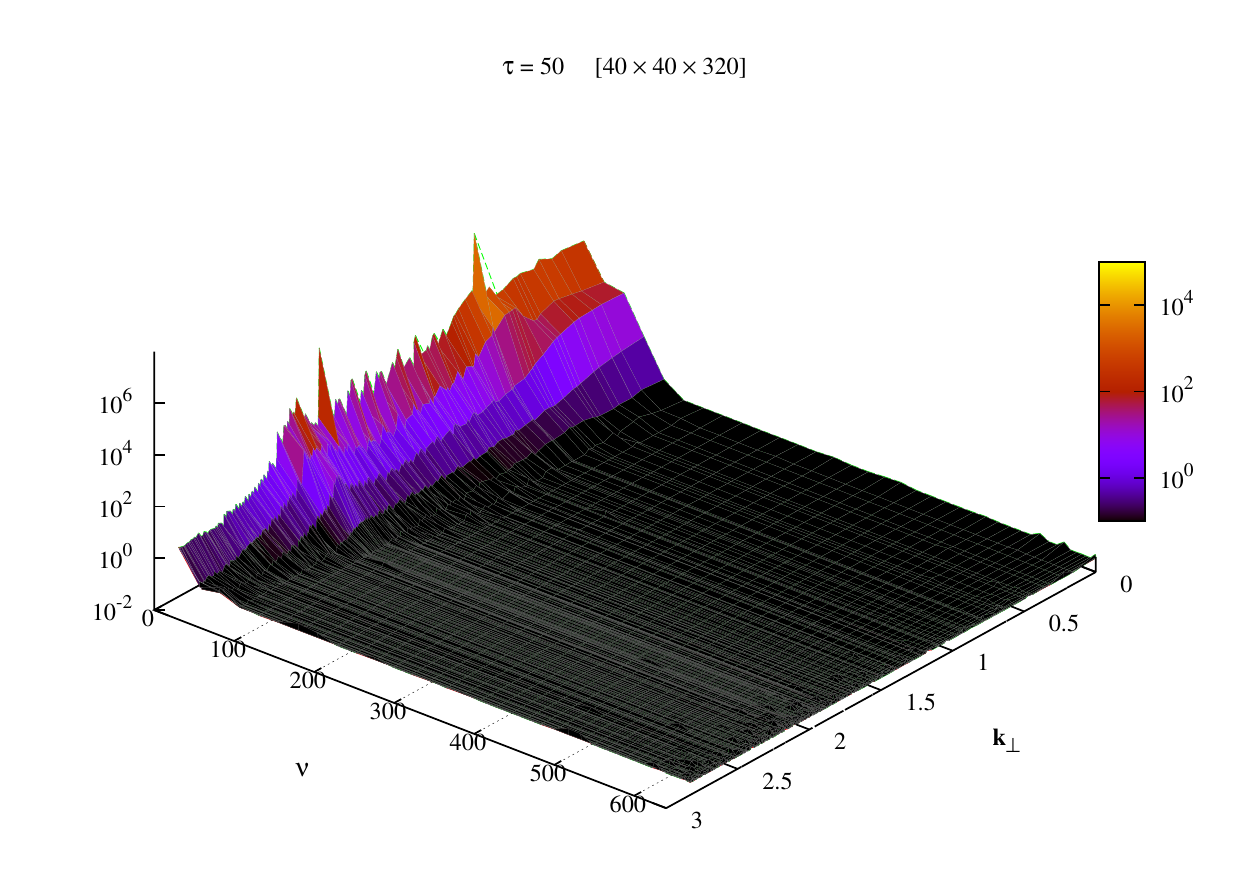}}}
\end{center}
\begin{center}
\resizebox*{7cm}{!}{{\includegraphics{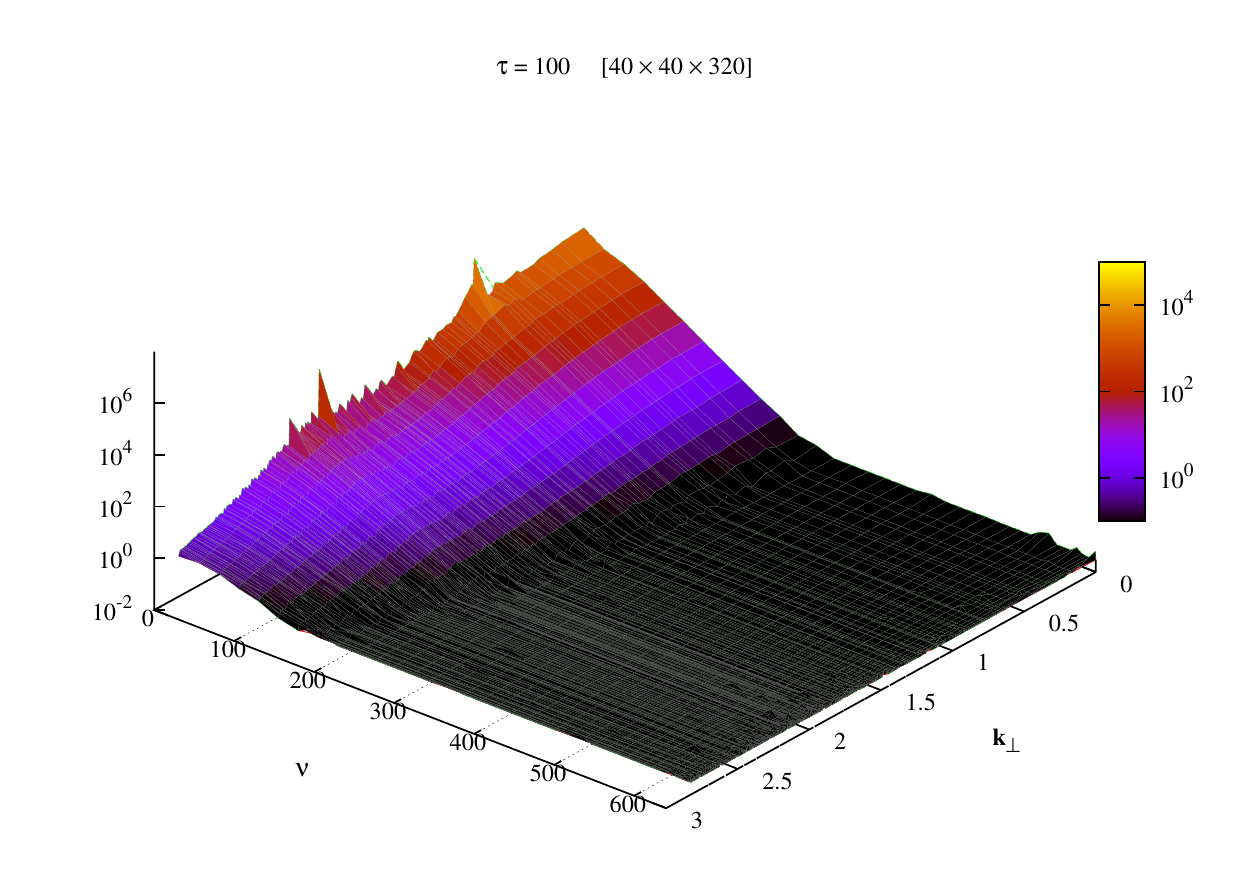}}}
\resizebox*{7cm}{!}{{\includegraphics{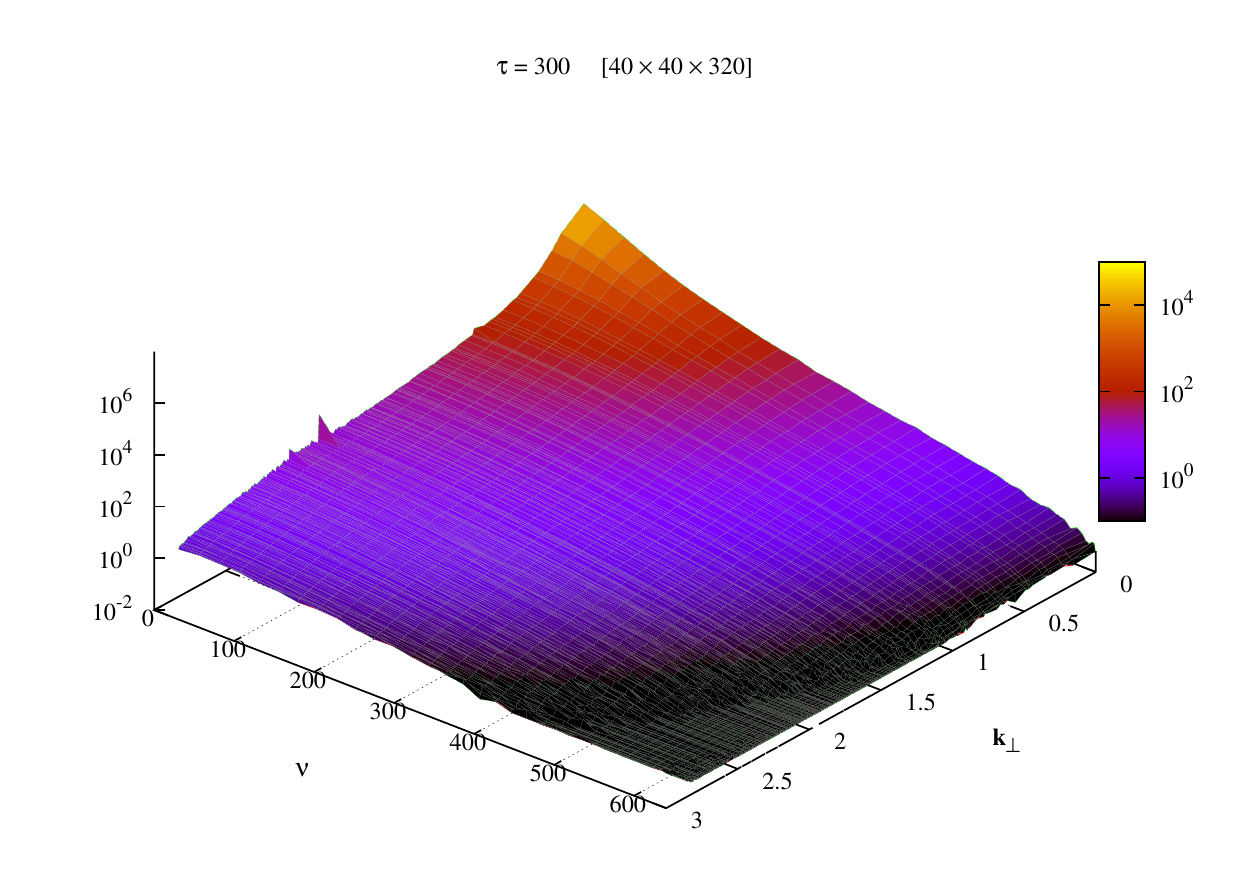}}}
\end{center}
\caption{\label{fig:fk320}Occupation number at times $\tau=10, 50, 100$
  and $300$, on a $40\times40\times
  320$ lattice.}
\end{figure}

A computationally expensive but otherwise straightforward way to avoid this
problem is to have a finer discretization in the rapidity direction, dividing the unit rapidity interval we are considering
in a large number $N$ of lattice spacings. As $N$ increases, we can
push the time evolution to later times. In the table \ref{tb:tmax}, we
give an estimate (obtained by computing the occupation number at
various times and observing when its support first reaches $\nu_{\rm
  max}$) of the maximal time, for a given
$N$, when lattice artifacts are negligible.
\begin{table}[htbp]
\begin{center}
\begin{tabular}{|r|c|}
\hline
$N$ & $\tau_{\rm max}$ \\
\hline
80 & 120 \\
\hline
160 & 220 \\
\hline
320 & 300 \\
\hline
\end{tabular}
\end{center}
\caption{\label{tb:tmax}Maximal time as a function of the number of
  lattice spacings in the $\eta$ direction.}
\end{table}
In the rest of the paper, we shall present results obtained on a
$40\times40\times 320$ lattice, which allows us to go safely up to
$\tau=300$. The occupation numbers at the times $\tau=10, 50, 100$ and
$300$ are shown in the figure \ref{fig:fk320}.

\section{Energy-momentum tensor} \label{sec:enertensor} 
\subsection{Definition}
We are now ready to explore the evolution of the energy-momentum
tensor. Besides the relaxation of the system towards an equation of
state, the longitudinal expansion means that there is no guarantee
that the transverse and longitudinal pressures will be equal. However,
for hydrodynamics to be applicable, the anisotropy of the pressure
tensor should not be large. It is therefore of the utmost importance
to determine whether the longitudinal and transverse pressures ever
approach each other for the expanding system.
 
In the $(\tau,\eta,\x_\perp)$ system of coordinates, the diagonal
components of the energy-momentum tensor are given by
\begin{eqnarray}
\varepsilon&=& \frac{1}{2}\Big(
\dot\phi^2+({\bs\nabla}_\perp\phi)^2+\tau^{-2}(\partial_\eta\phi)^2
\Big)+V(\phi)
\nonumber\\
T^{xx}&=& \frac{1}{2}\Big(
\dot\phi^2+(\partial_x\phi)^2-(\partial_y\phi)^2-\tau^{-2}(\partial_\eta\phi)^2
\Big)-V(\phi)
\nonumber\\
T^{yy}&=& \frac{1}{2}\Big(
\dot\phi^2-(\partial_x\phi)^2+(\partial_y\phi)^2-\tau^{-2}(\partial_\eta\phi)^2
\Big)-V(\phi)
\nonumber\\
\tau^2 T^{\eta\eta}&=& \frac{1}{2}\Big(
\dot\phi^2-(\partial_x\phi)^2-(\partial_y\phi)^2+\tau^{-2}(\partial_\eta\phi)^2
\Big)-V(\phi)\; .
\label{eq:pressures}
\end{eqnarray}
The transverse pressure is $P_{_T}\equiv(T^{xx}+T^{yy})/2$ and the
longitudinal pressure is $P_{_L}\equiv \tau^2 T^{\eta\eta}$. All these
quantities, written here for a single classical field configuration,
must of course be averaged over the Gaussian ensemble of initial
conditions derived in section \ref{sec:specfluc}. 

When computing the energy momentum tensor, it is important to do so on
a lattice that has a fine enough spacing in the rapidity direction, as
explained in the previous section. The artifacts one obtains at large
time if this is not the case are illustrated and discussed in the
appendix \ref{sec:artifacts}. In the rest of this section, all the
numerical results have been obtained on a $40\times40\times320$
lattice.

It is also important to subtract the contribution of the
vacuum. Indeed, our resummation leads to severe ultraviolet
divergences in the energy-momentum tensor that behave generically
like the fourth power of the ultraviolet cutoff (here, the inverse
lattice spacing). This unwanted term is a pure vacuum contribution
and it can be removed by subtracting the result from another
computation in which the background field $\varphi$ in
eq.~(\ref{eq:fluct1}) is set to zero (``vacuum''). In all the
quantities presented later on, this subtraction has been performed.

\subsection{Equation of state}
\begin{figure}[htbp]
\begin{center}
\resizebox*{10cm}{!}{\includegraphics{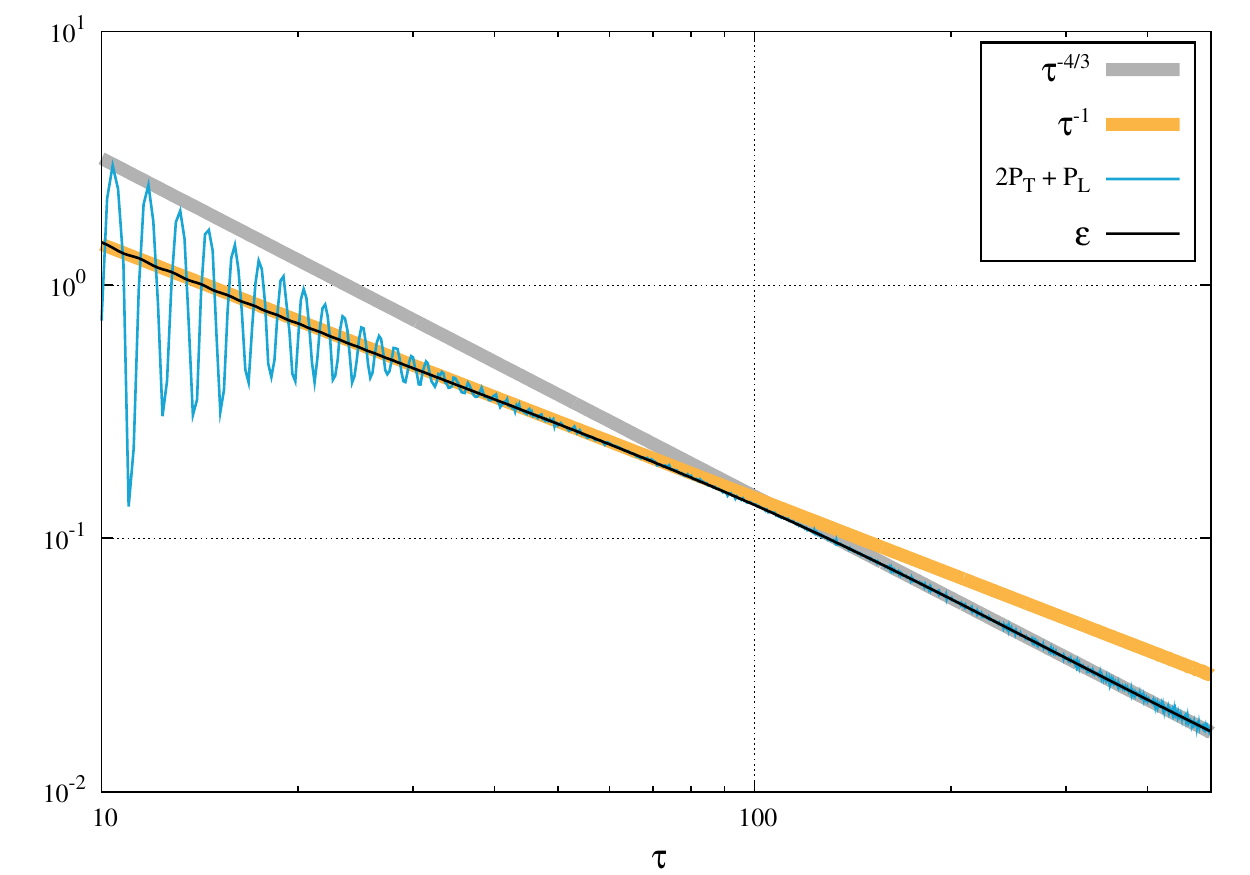}}
\end{center}
\caption{\label{fig:evol2} Time evolution of the energy density and of
  the trace of the pressure tensor, compared to the power law
  behaviors $\tau^{-4/3}$ and $\tau^{-1}$. Note: we are plotting the
  absolute value of the pressure, since at early times its sign
  changes periodically.}
\end{figure}

The first thing to assess is whether the system obeys an equation of
state. Towards that end, we plot in the figure \ref{fig:evol2}
the energy density and the sum of the three pressures.
As in a system with a fixed volume (see \cite{DusliEGV1}), the
pressure oscillates rapidly initially, with the oscillations being damped and disappearing eventually. When this happens, one sees that
\begin{equation}
\epsilon = 2P_{_T}+P_{_L}\; ,
\end{equation}
within statistical errors.

If one recalls the conservation equation\footnote{This is an exact
  equation, valid whether the system is in equilibrium or not. Note
  that since the energy density has a smooth derivative at $\tau=0^+$
  (see for instance the figure \ref{fig:checktau}), the longitudinal
  pressure must be negative and equal to $P_{_L}=-\epsilon$ at
  $\tau=0^+$. This is indeed the case but not obvious from the
  figures, where we plot the absolute value of the pressures.} that
drives the time evolution of the energy density,
\begin{equation}
\frac{\partial\epsilon}{\partial\tau}+\frac{\epsilon+P_{_L}}{\tau}=0\; ,
\end{equation}
it is also instructive to compare the time dependence of the energy
density with two power laws that have a special physical meaning:
\begin{itemize}
\item[{\bf i.}] $\tau^{-1}$, expected for a system whose longitudinal
  pressure is negligible,
\item[{\bf ii.}] $\tau^{-4/3}$, expected when $P_{_L}=P_{_T}=\epsilon/3$.
\end{itemize}
As one can see, at the beginning of the evolution, the energy density
decreases approximately like $\tau^{-1}$, while it is well fitted by
$\tau^{-4/3}$ at later times.  This is suggestive of the fact that the
system is initially highly anisotropic, and then becomes nearly
isotropic later on.

\subsection{Isotropization}
\begin{figure}[htbp]
\begin{center}
\resizebox*{10cm}{!}{\includegraphics{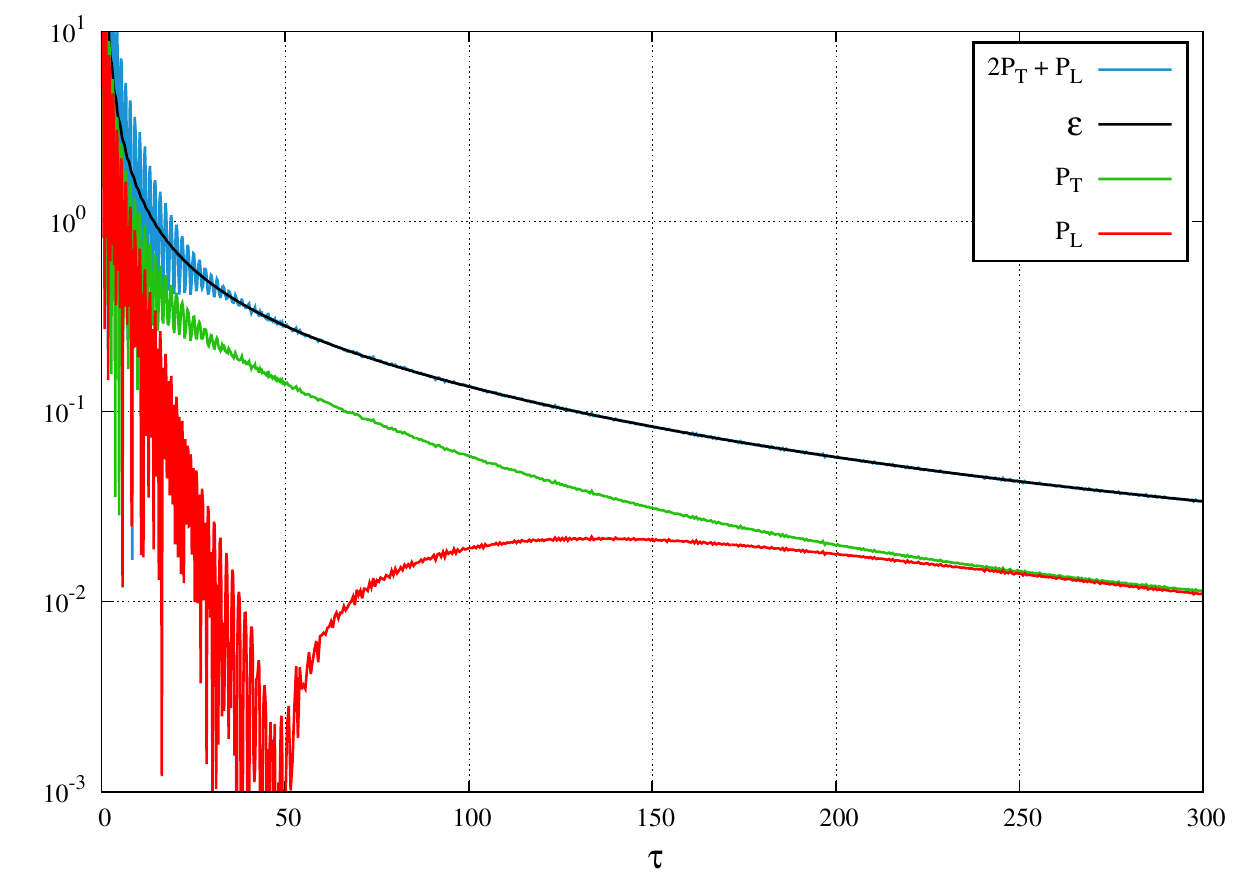}}
\end{center}
\caption{\label{fig:evolP} Time evolution of the diagonal components
  of $T^{\mu\nu}$, and of the trace of the pressure tensor. Note: we
  are plotting the absolute value of the pressures, since at early
  times their sign changes periodically.}
\end{figure}
Given the above results on the time dependence of the energy density,
let us now look at the evolution of the transverse and longitudinal
pressures separately. This is illustrated in figure
\ref{fig:evolP}.  (We also plot the energy density and the
sum of the three pressures in order to compare the timescales for the
relaxation of the pressures and for isotropization.)  In this plot,
one can distinguish three stages in the time evolution of the
pressures:
\begin{itemize}
\item[{\bf i.}]  At early times (on timescales comparable to that of
  the relaxation of the pressure or shorter), one sees that the
  magnitude of the longitudinal pressure drops very quickly. When well
  defined quasiparticles can be identified in the spectrum of the
  theory, this can be understood from the kinetic theory formula for
  the longitudinal pressure
  \begin{equation}
    P_{_L} = \int\frac{\d^3\p}{(2\pi)^3}\; \frac{p_z^2}{|\p|} \;f(\p)\;, 
  \end{equation}
  as a consequence of the red-shifting of the longitudinal momenta due
  to the expansion of the system. Indeed, particles with a non-zero
  $p_z$ eventually escape from the unit slice of rapidity whose
  evolution we are considering. This is illustrated in figure
  \ref{fig:expansion}.
\begin{figure}[htbp]
\begin{center}
\resizebox*{8.5cm}{!}{\includegraphics{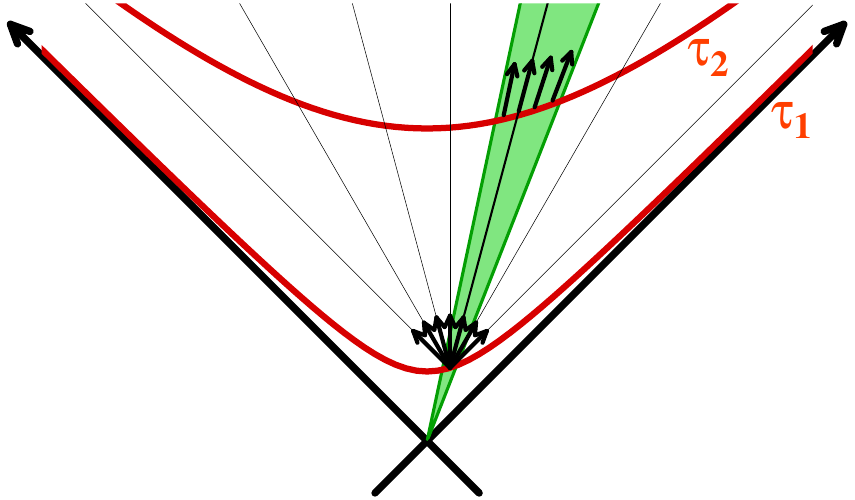}}
\end{center}
\caption{\label{fig:expansion} Evolution of the distribution of
  longitudinal momentum in a slice of rapidity, for particles
  undergoing free streaming. The thick arrows represent the velocities
  of the particles. The thin straight lines represent the
  trajectories of free particles--at constant velocity. At the
  time $\tau_1$, we assume that the rapidity slice under consideration
  contains particles with all longitudinal momenta. After some period
  of free streaming to the time $\tau_2$, only particles whose
  momentum rapidity $y$ equals the space-time rapidity $\eta$ are left.}
\end{figure}
At early times, the system we are considering is arguably not amenable
to a description in terms of quasiparticles. However, the above
argument about redshifting applies equally to the longitudinal
pressure of a system of fields, and implies a rapid decrease of its
magnitude.

\item[{\bf ii.}]  Next, a dramatic change of behavior occurs. The longitudinal pressure increases rapidly and approaches 
   the transverse pressure. This change begins when the
  oscillations of the pressure have become small. Moreover, from 
  figure \ref{fig:fk320}, we see that this occurs when the particle
  distribution starts to expand and occupy modes in the $\nu$ direction.

\item[{\bf iii.}]  Eventually, the longitudinal pressure becomes
  very close to the transverse one, albeit at times much later than the
  relaxation of the pressure. This confirms what was guessed on the
  basis of the time dependence of the energy density alone.
\end{itemize}

\begin{figure}[htbp]
\begin{center}
\resizebox*{10cm}{!}{\includegraphics{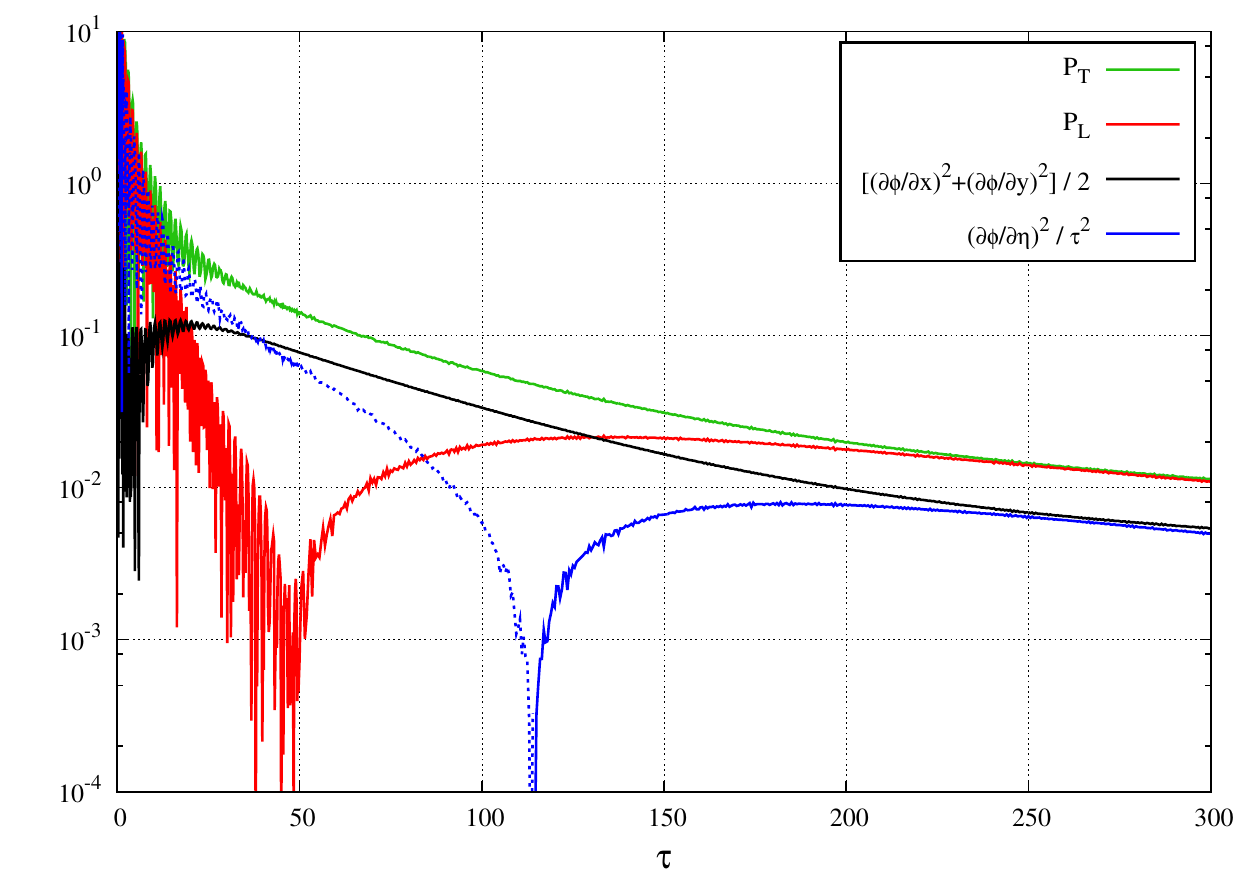}}
\end{center}
\caption{\label{fig:evol} Time evolution of the transverse and
  longitudinal gradients that enter in the pressures. The dotted
  line means that the corresponding quantity is negative.  (We
  have plotted here its absolute value.)  We also plot the time evolution
  of the transverse and longitudinal pressures for easy reference.}
\end{figure}

The difference between the transverse and longitudinal pressures
essentially arises from the sign with which the various spatial
gradients of the field enter in the formulas (\ref{eq:pressures}).  In
particular, it easy to see from these formulas that the two pressures are
equal provided that
\begin{equation}
\frac{1}{2}({\bs\nabla}_\perp\phi)^2 = \frac{1}{\tau^{2}}(\partial_\eta\phi)^2\; .
\end{equation}
In order to better understand the evolution of the two pressures, we
have represented in figure \ref{fig:evol} the expectation values
of the two quantities $({\bs\nabla}_\perp\phi)^2/2$ and
$\tau^{-2}(\partial_\eta\phi)^2$.  Note that these quantities have
been vacuum-subtracted, as explained after eq.~(\ref{eq:pressures}),
which means that they are not necessarily positive. Negative values
are indicated in the plot by a dotted line instead of a solid one.

At very early times, the transverse gradient is moderately small and
positive, while the longitudinal gradient is large and negative (after
we have subtracted the contribution from the vacuum).  Up to
$\tau\approx 10$, the transverse gradient remains roughly constant
(more precisely, it oscillates around a constant value), while the
magnitude of the longitudinal gradient decreases very rapidly. (Note
that since it is negative, this means that it is in fact increasing with
time.)

For a short period of time after $\tau\approx 10$, the
transverse gradient increases mildly. From the plots showing the
evolution of the occupation number in the figure \ref{fig:fk320}, this
is clearly related to the expansion of the particle distribution in
$\k_\perp$. During this period, the transverse gradient is much larger
than the longitudinal one (which is still negative), and therefore the
longitudinal pressure is much smaller than the transverse one.

A qualitative change in behavior occurs around $\tau\approx 50$: the
longitudinal gradient evolves faster than the transverse one and it
increases rapidly, approaching zero to become positive shortly after
$\tau\approx 100$. The consequence on the longitudinal pressure is a
rapid increase despite the expansion of the system. Microscopically
the increase of the longitudinal gradients is presumably due to the
resonance that can reshuffle efficiently momenta now that the
expansion rate of the system is lower. (Before that point, any particle
that was scattered at a non-zero $p_z$  immediately escaped from
the rapidity slice under consideration.) This trend continues until
the transverse and longitudinal gradients become comparable, at which
point the two pressures are nearly equal.

In appendix \ref{app:iso}, we study how the isotropization
timescale varies when we change the amplitude of the background field
and the coupling strength; our results show that isotropization is
faster for larger fields and/or larger couplings.

\section{Comparison with hydrodynamics}
\label{sec:hydro}
The previous results indicate that the trace of the pressure tensor
relaxes to its equilibrium value. Subsequently, the longitudinal
pressure approaches the transverse one and the system becomes
isotropic despite the continuing longitudinal expansion. A natural question to
ask is whether the details of this time evolution are compatible with
hydrodynamics\footnote{See \cite{Teane1} for a recent review on
  relativistic hydrodynamics. One can also find an interesting
  comparison between an exactly solvable (via the AdS/CFT
  correspondence) toy model at strong coupling and relativistic
  hydrodynamics in ref.~\cite{HelleJW1}.}.  Of course it does not
make sense to try answering this question before the time at which the
pressures become well defined, namely, before the pressure oscillations
have disappeared. Further, it is clear that to map our pressures onto
hydrodynamics, a minimal requirement is to include a shear
viscosity that will allow the pressure tensor to be anisotropic. Given the geometry of our setup, it is 
sufficient to consider boost invariant hydrodynamics. Also, since our
problem is translationally invariant in the transverse directions, we can
set the transverse fluid velocity to zero.

\subsection{Energy and momentum conservation}
\begin{figure}[htbp]
\begin{center}
\resizebox*{10cm}{!}{\includegraphics{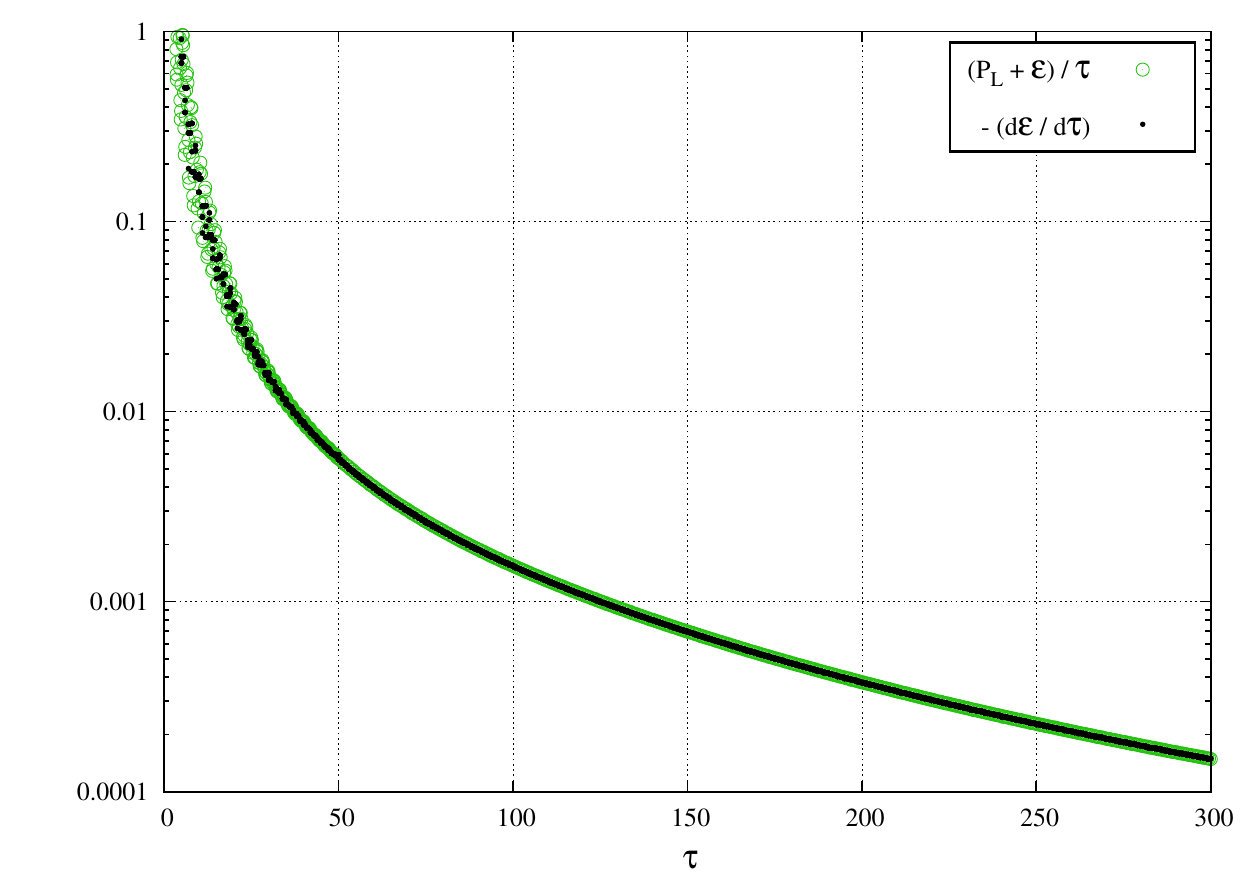}}
\end{center}
\caption{\label{fig:depsilon} Comparison between the two sides of
  eq.~(\ref{eq:hydro1}).}
\end{figure}

In this situation, the time evolution of the energy density is related
to the longitudinal pressure, 
\begin{equation}
\frac{\partial\epsilon}{\partial\tau}
=-\frac{\epsilon+P_{_L}}{\tau}\; .
\label{eq:hydro1}
\end{equation}
A first consistency check that we can make is to verify that this
equation holds from our numerical results. This is shown in fig. 
\ref{fig:depsilon}, where one clearly sees that this equation is
satisfied.
This should of course not be a surprise because the equation comes
directly from the conservation of energy and momentum and should be
satisfied regardless of whether hydrodynamics applies or not.

\subsection{Comparison with first order viscous hydrodynamics}
The previous check was merely a verification that our numerical
approach fulfills basic conservation laws. In order to go further, let
us recall that in first order viscous Bjorken hydrodynamics the
longitudinal and transverse pressures would be given by
\begin{eqnarray}
P_{_T}&=& \frac{\epsilon}{3}+\frac{2\eta}{3\tau}\nonumber\\
P_{_L}&=& \frac{\epsilon}{3}-\frac{4\eta}{3\tau}\; ,
\label{eq:visc-P}
\end{eqnarray}
where $\eta$ is the shear viscosity. This ansatz assumes that we are
already in the regime where $\epsilon = 2P_{_T}+P_{_L}$, and
attributes the difference between the two pressures to the effect of
the shear. 

\begin{figure}[htbp]
\begin{center}
\resizebox*{10cm}{!}{\includegraphics{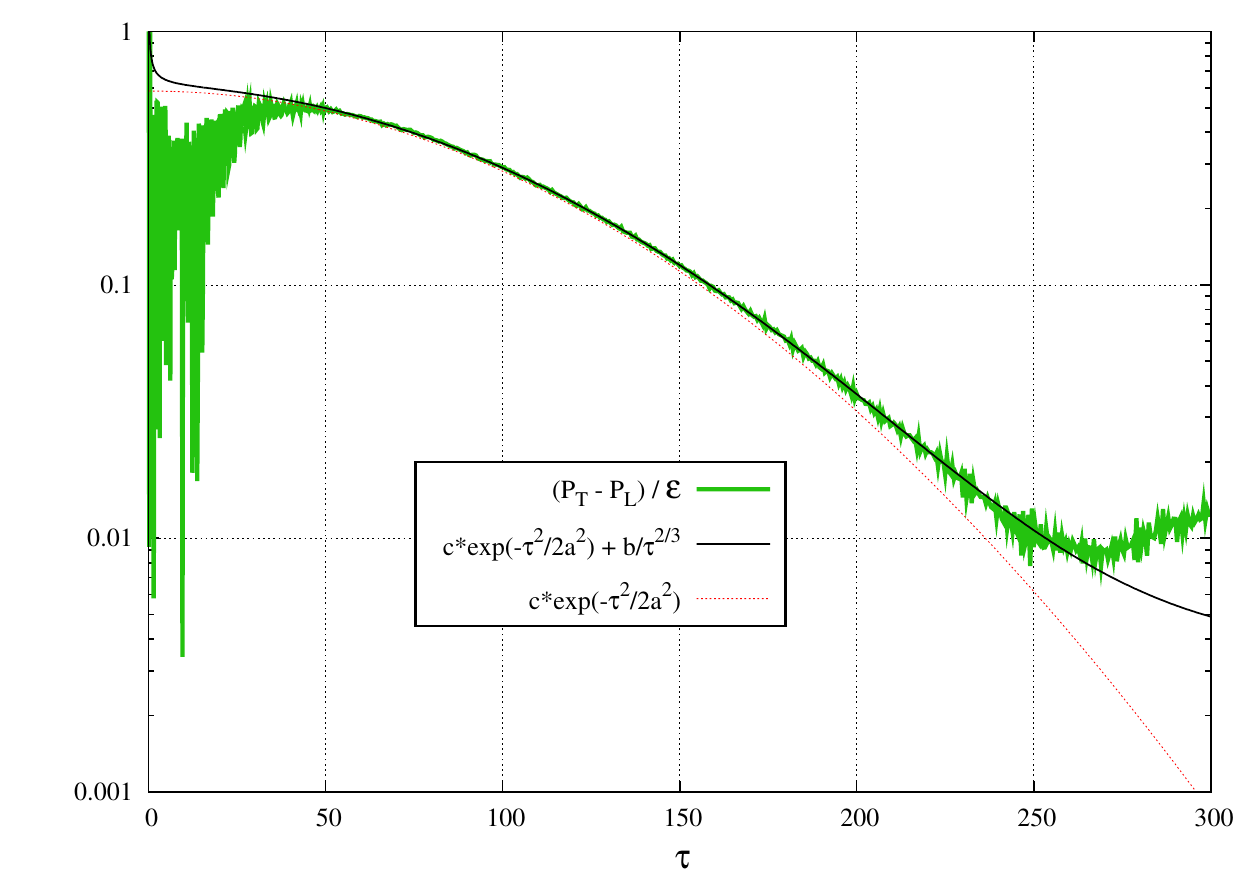}}
\end{center}
\caption{\label{fig:dP} Time evolution of the pressure anisotropy
  $(P_{_T}-P_{_L})/\epsilon$. Red dotted line~: fit in
  $\exp(-\tau^2/2a^2)$. Black solid line~: fit in $\exp(-\tau^2/2a^2)$
  plus a power law. The relaxation time in the exponential is
  $a\approx 82.9$ lattice spacings, and the coefficient of the power
  law correction is $b\approx 0.182$ (in lattice units).}
\end{figure}

First of all, we can compare the time evolution of the pressure
anisotropy, $(P_{_T}-P_{_L})/\epsilon$, shown in fig.~\ref{fig:dP}, with what one would expect in a hydrodynamical expansion.
From the time where the oscillations of the pressures have died out to
the time where this anisotropy has become very small, it can be fitted
by an exponential form $\exp(-\tau^2/2a^2)$. At later times, the quality
of the fit can be improved slightly by adding a term in
$\tau^{-2/3}$. (Note that at times greater than $\tau\gtrsim 280$, the
anisotropy rises again due to lattice artifacts.)  

If the decrease of the anisotropy was driven by viscous hydrodynamics,
one would expect its behavior to be a power law $\tau^{-2/3}$, not an
exponential.  Although it is not easy to pinpoint what precisely is
causing this exponential falloff of the anisotropy, it seems very
plausible that the presence of instabilities in the microscopic
dynamics of the system is responsible. One may argue that the
hydrodynamical regime only starts when the power law term in
$\tau^{-2/3}$ becomes important--roughly after $\tau\sim200$
lattice spacings. Then one could turn the coefficient $b$ of the
$\tau^{-2/3}$ term into a value of the ratio $\eta/s$. The
dimensionless ratio $\eta/s$ determines the strength of the viscous
effects on the hydrodynamical evolution of the system. In a scale
invariant theory like the one we consider here, it is natural that this
ratio is a fixed number\footnote{This statement is true at the order
  at which we analyze the system but would be violated once one 
  includes running coupling effects.}  that depends only on the value
of the coupling constant $g$. In the hydrodynamical regime, one would
have
\begin{equation}
\left[\frac{P_{_T}-P_{_L}}{\epsilon}\right]_{\rm hydro}
=2\,\frac{\eta}{s}\,\frac{s}{\tau\epsilon}\approx 
\underbrace{\frac{\eta}{s}\,\frac{2}{A^{1/4}}}_{b}\;\frac{1}{\tau^{2/3}}\; ,
\end{equation}
where we have used the Stefan-Boltzmann formulas to estimate the energy and entropy density (assuming the system is close to
equilibrium),
\begin{equation}
\epsilon = \frac{\pi^2 T^4}{30}\quad ,\;
s = \frac{2\pi^2 T^3}{45}\quad,\;
s \approx \epsilon^{3/4}\; .
\label{eq:s}
\end{equation}
and where $A$ is the coefficient in the asymptotic behavior of the
energy density, $\epsilon\approx A \tau^{-4/3}$. From this formula, we
extract the following value for the ratio $\eta/s$,
\begin{equation}
\frac{\eta}{s}\approx 0.26\; , 
\label{eq:visco1}
\end{equation}
with a large systematic uncertainty due to the limitations of our analysis outlined previously. 

To go beyond this simple comparison, we should solve
eq.~(\ref{eq:hydro1}) with the ansatz of eq.~(\ref{eq:visc-P}) for the
longitudinal pressure. To close the equation, we need a way
to relate $\eta$ to $\epsilon$, with eq.~(\ref{eq:s}).  Even without
solving the hydrodynamical equations, we can use this approximation
for the entropy density in order to extract an effective $\eta/s$ from
our numerical results. This amounts to rewriting
eqs.~(\ref{eq:visc-P}) as
\begin{equation}
P_{_T} = \frac{\epsilon}{3} 
+ \frac{2}{3\tau}\left[\frac{\eta}{s}\right]_{\rm eff}
\epsilon^{3/4}
\quad,\quad
P_{_L} = \frac{\epsilon}{3} 
- \frac{4}{3\tau}\left[\frac{\eta}{s}\right]_{\rm eff}
\epsilon^{3/4}\; .
\end{equation}
\begin{figure}[htbp]
\begin{center}
\resizebox*{10cm}{!}{\includegraphics{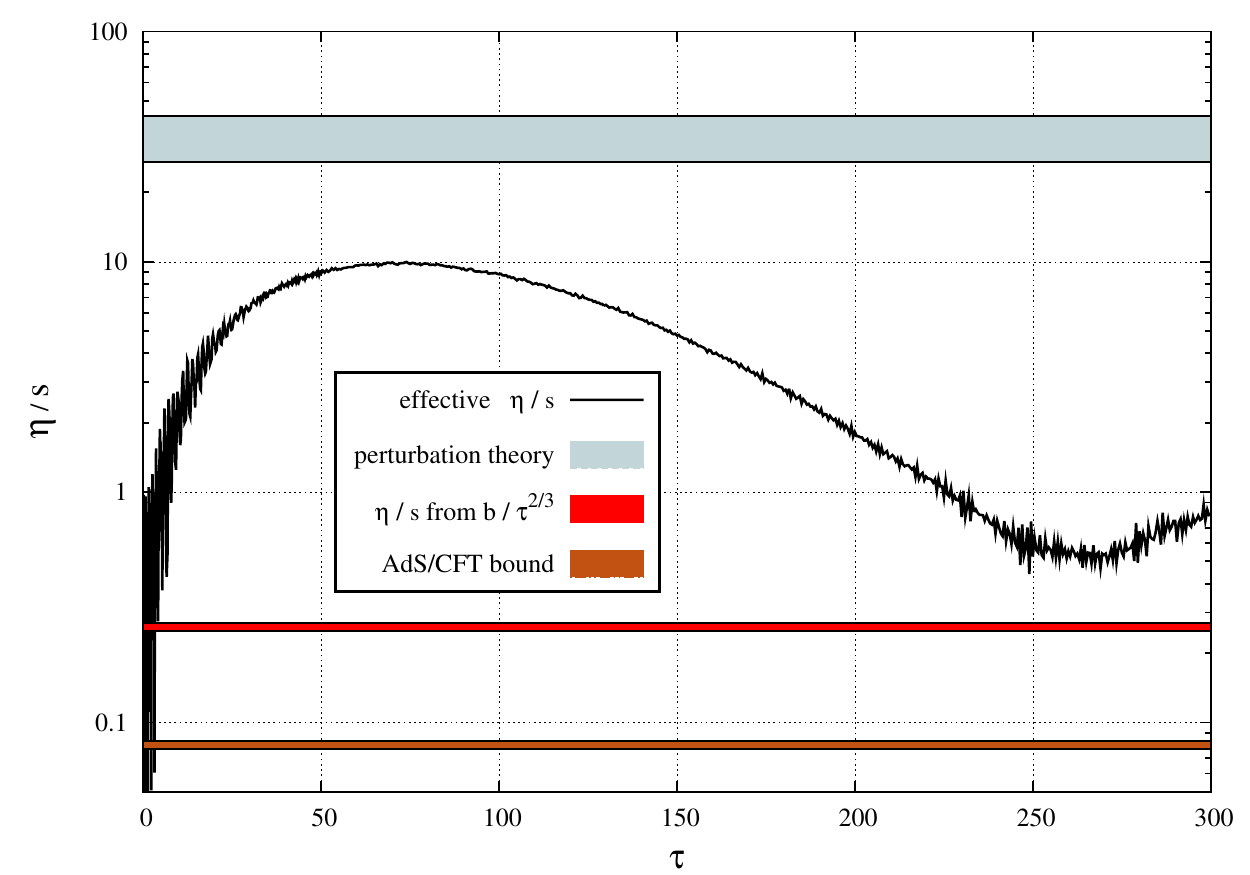}}
\end{center}
\caption{\label{fig:visco} Evolution of the effective $\eta/s$
  ratio. We have also represented the perturbative value of this
  ratio, the value extracted from the power term
  (eq.~(\ref{eq:visco1})), and the conjectured lower bound $1/4\pi$ derived in
  AdS/CFT at strong coupling.}
\end{figure}
Contrary to the estimate in eq.~(\ref{eq:visco1}), where we attributed
to the viscosity only the power law term in the difference
$P_{_T}-P_{_L}$, we now define this effective viscosity from the full
difference, including also the exponential term. This effective ratio
can be computed at each time, and is shown in the figure
\ref{fig:visco}. The first thing one sees is that it is not constant,
which certainly means that there are discrepancies with
hydrodynamics. The value of this ratio before $\tau\approx 70$
probably does not make much sense, because the pressures are
oscillating before this time. After that, the ratio starts near a
value of $10$ and decreases rapidly to reach a value of about $0.5$.
In addition to the non-constancy of this ratio, it should be noted
that its value is quite small compared to the perturbative value. For
a real scalar field theory with a $g^2\phi^4/4!$ coupling, the lowest
order perturbative result~\cite{Jeon2} is roughly
\begin{equation}
\frac{\eta}{s}\sim \frac{10^4}{g^4}\; .
\end{equation}
Given that $g=4$ in our simulation, one would expect $\eta/s\sim 40$,
which is much larger than the effective ratio we observe. In other
words, the system of fields we have studied is a much better fluid
than one would expect on the basis of the perturbative estimates of the
viscosity to entropy density ratio. This anomalously small value of the $\eta/s$ ratio is perhaps 
related to the phenomenon discussed in \cite{AsakaBM1,AsakaBM2} where
it is argued that, in systems subject to turbulent unstable fields,
momentum transport may occur as if the viscosity were much smaller
than naive estimates from transport cross-sections.

To carry the comparison of our simulations with hydrodynamics even further, we must solve the
hydrodynamic equations; using eq.~(\ref{eq:s}), the evolution of
$\epsilon$ in viscous hydrodynamics has the closed form 
\begin{equation}
\frac{\partial\epsilon}{\partial\tau}
+\frac{4}{3}\frac{\epsilon}{\tau}
-\frac{4\eta}{3s}\,\frac{\epsilon^{3/4}}{\tau^2}
=0\; .
\label{eq:hydro2}
\end{equation}
Comparisons between the two frameworks can be performed as follows:
\begin{itemize}
\item[{\bf i.}] Choose an initial time $\tau_0$, where hydrodynamics is initiated. This time should be well after the
  oscillations in the pressures have disappeared.
\item[{\bf ii.}] Initialize the energy density so that it has the same
  value in the two frameworks at $\tau_0$.
\item[{\bf iii.}] Set the value of the ratio $\eta/s$ in order to have
  the right values for the transverse and longitudinal pressures at
  $\tau_0$. This is always possible if $\tau_0$ is in the region where
  $\epsilon=2P_{_T}+P_{_L}$.
\item[{\bf iv.}] Solve (numerically) the differential equation
  (\ref{eq:hydro2}) from these initial conditions and compare with
  the evolution one obtains in classical statistical field theory.
\end{itemize}
\begin{figure}[htbp]
\begin{center}
\resizebox*{10cm}{!}{\includegraphics{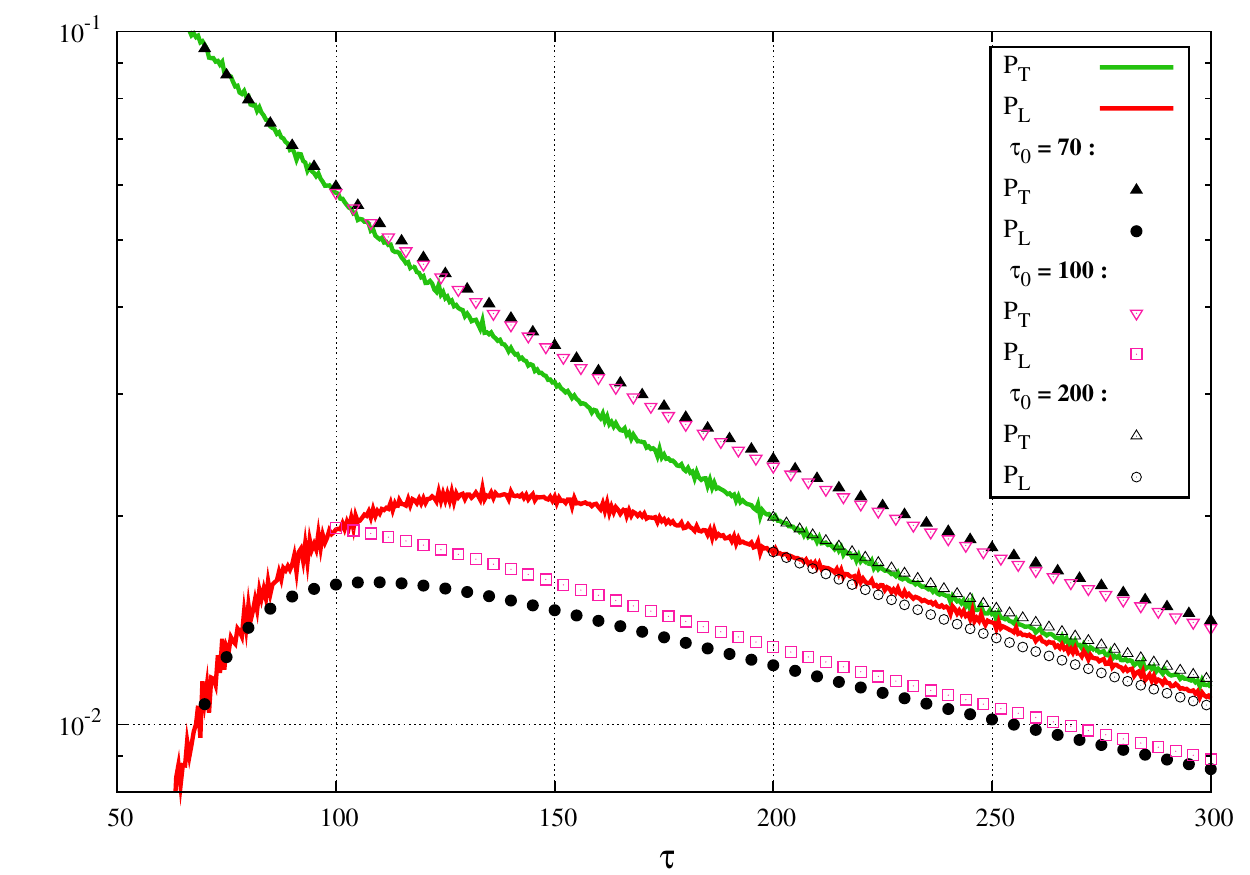}}
\end{center}
\caption{\label{fig:Hcomp} Comparison between classical statistical
  field theory and first order viscous hydrodynamics, for various
  starting times of the hydrodynamical evolution ($\tau_0 = 70, 100,
  200$). Solid lines: pressures in classical statistical field
  theory. Dots: results of first order viscous hydrodynamics.}
\end{figure}
The result of this comparison is shown in fig. \ref{fig:Hcomp},
for three values of the initial time $\tau_0$ for hydrodynamics. The
main lesson from this figure is that the pressure tensor isotropizes
much faster in classical statistical field theory than it does in
hydrodynamics. Hydrodynamics begun at earlier times cannot compete and thus flow and viscosity bounds 
derived from such exercises are suspect. 

The last pair of hydro curves, with an initial time $\tau_0=200$,
remains close to the true evolution, but this is only because it {\sl
  starts} with an already small amount of anisotropy. Therefore, for hydrodynamics to describe accurately the evolution, 
 hydrodynamical evolution should be initiated at a time where the
pressure tensor is already close to isotropy. Our 
comparison is based on first order viscous hydrodynamics
only. Arguably, viscous corrections are large in the system when
$P_{_T}$ and $P_{_L}$ differ substantially and it is unclear whether
 first order hydrodynamics is a valid approximation.

We close this section with a generic comment about the
isotropization time. The precise value of the time we obtain in
this model of scalar fields is of little relevance for the analogous 
question in QCD because the two theories are quite different in many
respects. However, we expect the conclusion that the true
isotropization time is considerably shorter than the time one would
infer from hydrodynamics (or equivalently from a Boltzmann equation)
is quite generic in theories with instabilities.  The other lesson from 
this comparison is that such systems may have an effective viscosity
to entropy density ratio which is much smaller than what one expects
from perturbation theory. Therefore one could have nearly ideal
fluid behavior without invoking strong coupling. Moreover, we know
from \cite{EpelbG1} that full thermalization of this system takes
much longer than the times considered here. We may conclude
that full thermalization is not necessary for a hydrodynamical
description; only near isotropy is important.

\section{Summary and outlook} 
\label{sec:conclusion}

This work extends our previous study of the role of quantum
fluctuations on the thermalization of strong fields (subject to
instabilities) to the situation where the system undergoes
longitudinal expansion.  We considered a one-component scalar field
theory, which is much simpler to study numerically than Yang-Mills
theory, and implemented the same resummation program that one would
employ in studying the early stages of heavy ion collisions in the
Color Glass Condensate framework.

As in the case of a system confined to a fixed box, the trace
of the pressure tensor exhibits rapid temporal oscillations that are quickly
damped via the phenomenon of phase decoherence.  Following this state we find
that the sum of
the three pressures become equal to
the energy density (a well-define equation of state exists), but the transverse
and longitudinal pressures remain different until full isotropization at an
even later time.  

The main new result of this study is that the pressure tensor
eventually becomes isotropic despite the longitudinal expansion of
the system.  After an initial decrease due to the rapid expansion at
early times, the longitudinal pressure increases at later times to
reach a value that is very close to that of the transverse
pressure. This goes a long way towards justifying the validity of the
hydrodynamical description for expanding systems, as encountered in
heavy ion collisions.

We also studied the time evolution of the particle
distribution. Even with a very far from equilibrium initial condition
that has a single $\k_\perp$ mode at $\tau=0^+$, a continuum of modes
rapidly get filled by the nonlinear interactions of the fields.
This study of the occupation number highlights an important
limitation of numerical simulations that use rapidity as the
longitudinal coordinate, namely, the particle distribution expands in $\nu$
--the conjugate momentum of rapidity-- over time and eventually
reaches the ultraviolet cutoff imposed on this variable by the lattice
spacing. Therefore in order to pursue the study until sufficiently
large times without being affected by these lattice artifacts, it is
necessary to use a lattice that has a very fine mesh in the rapidity
direction. This issue is by no means specific to a scalar theory, and
will be present when this approach is applied to Yang-Mills theory.

In all the plots presented in this paper, all dimensionful quantities
were expressed in units of the appropriate power of the transverse
lattice spacing. Since the $\phi^4$ scalar theory in $3+1$ dimensions
is scale invariant at the classical level (as is Yang-Mills theory),
one can re-express all of them in terms of a single dimensionful
physical parameter. In the Color Glass Condensate framework, this
parameter would be the saturation momentum $Q_s$ of the colliding
nuclei. Doing this for QCD would provide an answer to the following
fundamental question: {\sl how long does isotropization take, given
  the value of the saturation momentum and of the strong coupling
  constant?} However, the lessons of this exercise in scalar field
theory cannot be simply applied to the gauge theory case because the
former differs from QCD in a number of crucial ways. To list a few,
(i) the field content of QCD is much richer, (ii) there is no
meaningful way to compare the coupling constants of the two theories,
(iii) the strength of their instabilities are quite different, both in
their range and in their growth rate.

Thus it is best to see the present work as a proof of the concept that
instabilities can isotropize the pressures of an expanding system. The
next step obviously is to apply the same treatment to Yang-Mills
theory. In a previous work \cite{DusliGV1}, we derived the spectrum of
initial field fluctuations that one must use in this computation. The
numerical implementation of this approach to a Yang-Mills theory is
under way but will require a significantly larger computational effort
than what was needed for the present scalar theory.

\section*{Acknowledgements}
We would like to thank J.~Berges, J.-P.~Blaizot, J.~Liao, L.~McLerran,
and S.~Schlichting for useful discussions. FG and TE are supported by
the Agence Nationale de la Recherche project \#~11-BS04-015-01. FG
would also like to thank the Brookhaven National Laboratory and the
Institute of Nuclear Theory, where parts of this work were performed,
for their hospitality and support. K.D.  and R.V are supported by the
US Department of Energy under DOE Contract Nos.  DE-FG02-03ER41260 and
DE-AC02-98CH10886 respectively.

\appendix

\section{Lattice artifacts at late times}
\label{sec:artifacts}
As we have already explained in section \ref{sec:fk-evol}, the time extent of
the simulation is limited by the finite size of the lattice.  We demonstrated
that as the time increases the particle distribution expands to larger $\nu$ values,
($\nu$ is the momentum conjugate to rapidity), until it is artificially cut-off
by the discretization in rapidity.  

\begin{figure}[htbp]
\begin{center}
\resizebox*{10cm}{!}{\includegraphics{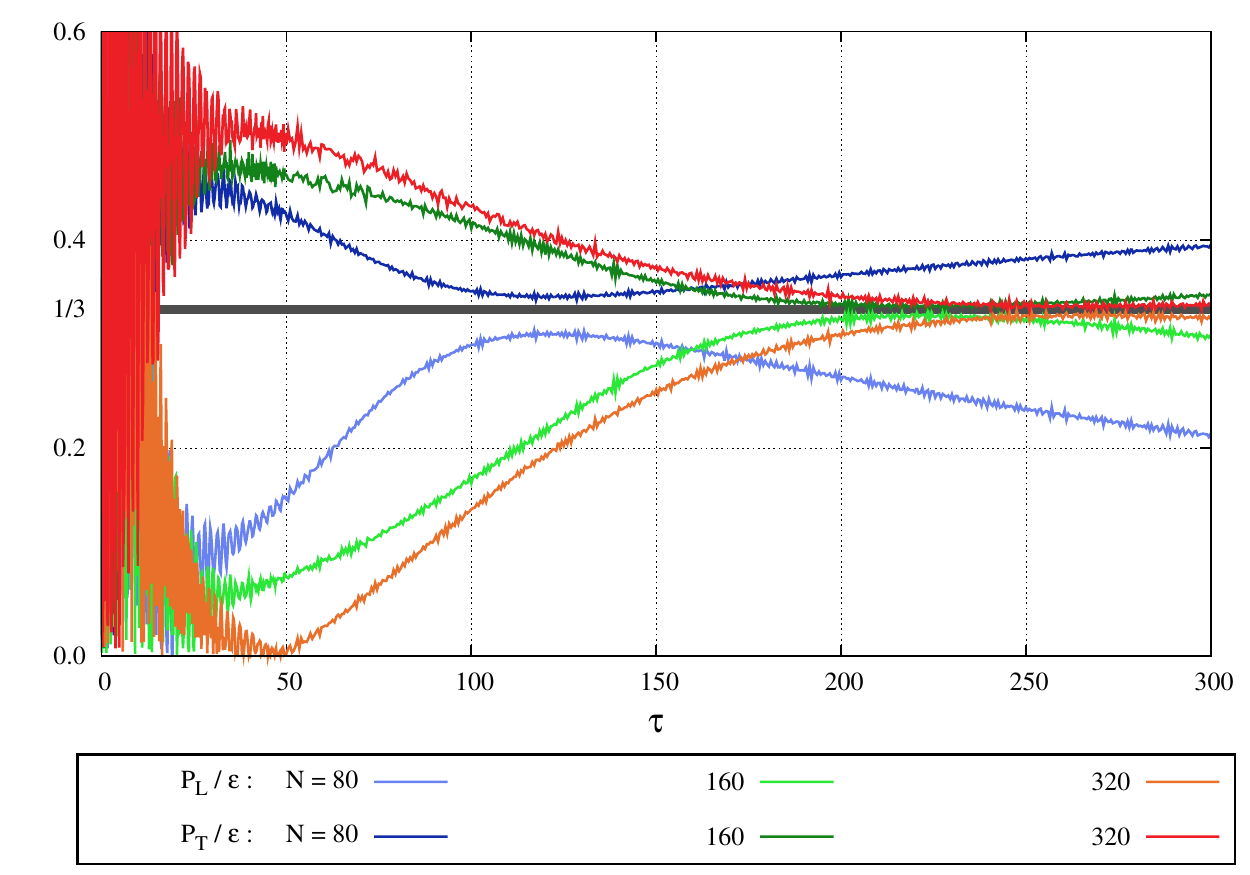}}
\end{center}
\caption{\label{fig:anis}Time evolution of the ratios of longitudinal
  and transverse pressure to the energy density, for several
  choices of discretization in rapidity.}
\end{figure}

In figure \ref{fig:anis}, we show the practical significance of this
effect on the numerical computation of the components of the
energy-momentum tensor, for lattices with increasing number of
longitudinal spacings $N$ (all describing a unit slice of rapidity).  We
see that when the number of lattice spacings in the rapidity
direction is small (see the curves for the $40\times 40\times 80$
lattice), the ratio $P_{_L}/\epsilon$ approaches the isotropic value
$1/3$ and then departs from this value. Likewise, the ratio
$P_{_T}/\epsilon$ approaches $1/3$ from above, before deviating from
this value (in such a way that $2P_{_T}+P_{_L}$ remains equal to
$\epsilon$).  The time at which this occurs agrees with the time when the
particle distribution reaches the longitudinal momentum cutoff as shown in
table~\ref{tb:tmax}.  Figure \ref{fig:anis} demonstrates that
this artifact appears at later times as the number of lattice spacings in
rapidity is increased.  

The additional complication of renormalization must also be considered when simulating scalar fields on a lattice.  While there is no mass term in our bare
Lagrangian, vacuum fluctuations induce a non--zero mass.  In a scalar $\phi^4$
theory, the leading contribution stems from the tadpole
graph having a quadratic UV divergence.
\begin{figure}[htbp]
\begin{center}
\resizebox*{10cm}{!}{\includegraphics{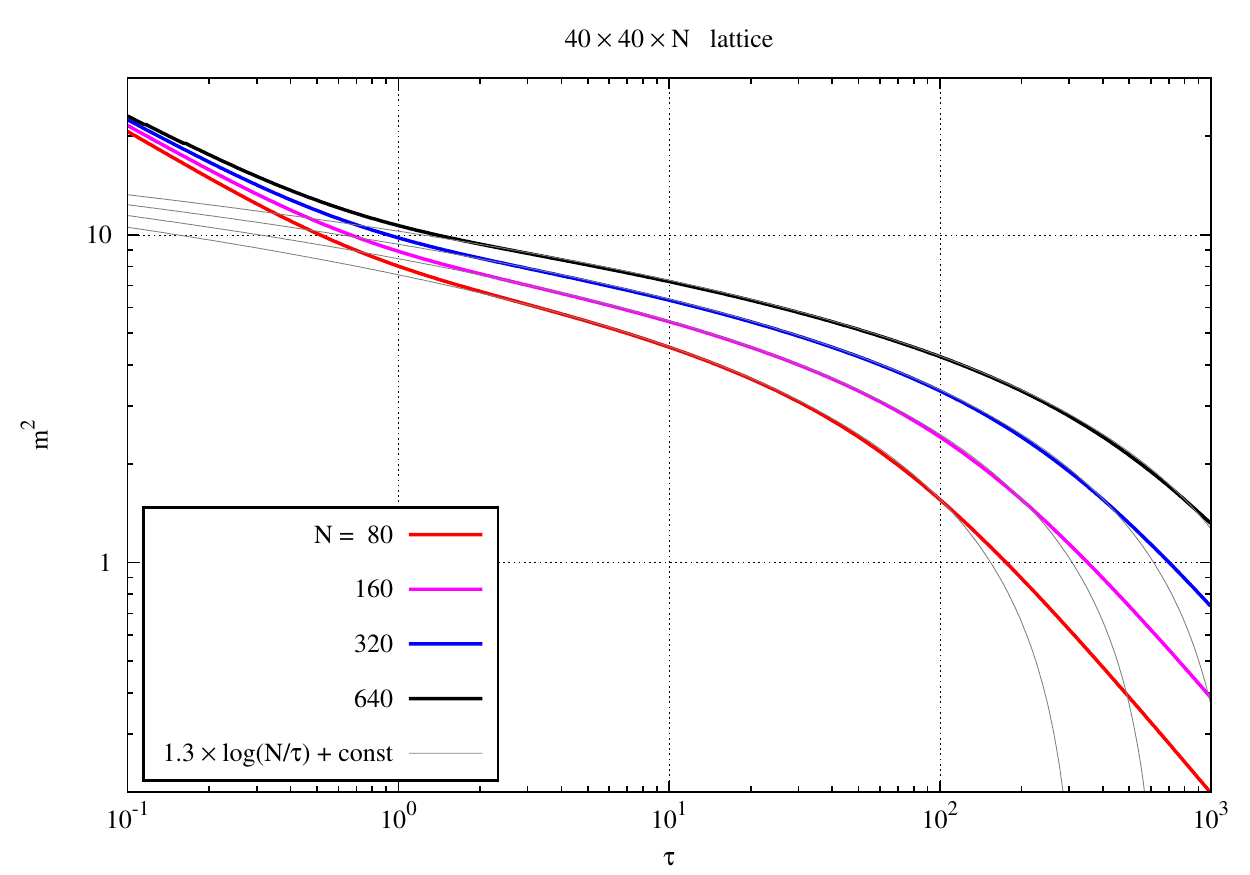}}
\end{center}
\caption{\label{fig:mass}Tadpole contribution to the mass as a
  function of time, for various longitudinal lattice spacings (and a
  coupling constant $g=4$).}
\end{figure}
The value of this tadpole is given as
\begin{equation}
m^2 = \frac{g^2\pi}{8}\int \frac{\d^2\k_\perp}{(2\pi)^2}\frac{\d\nu}{2\pi}\;e^{\pi\nu}\,
\left|H_{i\nu}^{(2)}(k_\perp\tau)\right|^2\; .
\end{equation}
which clearly exhibits the nature of its divergence.  On the lattice the above
integral is replaced by a discrete sum having a UV cutoff dictated by the
size of the lattice spacing.  In $\tau$--$\eta$ coordinates this mass is $\tau$
dependent (a fixed UV cutoff in $\nu$ corresponds to a time dependent cutoff in
$k_z\sim \nu/\tau$).  Figure \ref{fig:mass} shows the time evolution of $m^2$ for
various discretizations in rapidity.  At intermediate times the mass has a
logarithmic sensitivity to the lattice spacing.

In practice this means that computations done with the same bare
Lagrangian but having different longitudinal discretizations
correspond to different renormalized theories.  The logarithmic
dependence of this mass with $N$ means that it varies more rapidly at
small $N$ than at large $N$, something that can be seen in the figure
\ref{fig:anis} by comparing the changes from $N=80$ to $N=160$ and
from $N=160$ to $N=320$. In the present work, we have not attempted to
renormalize the bare parameters as this will be further complicated by
the fact that we are treating the transverse and longitudinal
coordinates on different footings.  Therefore, one should keep in mind
that the theory for which results are presented in the paper is not a
truly massless theory, and the comparison between computations done at
varying lattice spacings can only be made qualitatively.  We should
stress that this issue should not arise in gauge theories.  In this
case, the gauge invariance of the Wilson action prevents the generation of
a mass term for the gluons.

\section{How generic is it? Varying $g^2$ and $\varphi$}
\label{app:iso}
A natural question that arises is whether the isotropization that we
have observed is a generic phenomenon that occurs for all choices of
coupling strengths and background fields. 

\begin{figure}[htbp]
\begin{center}
\resizebox*{10cm}{!}{\includegraphics{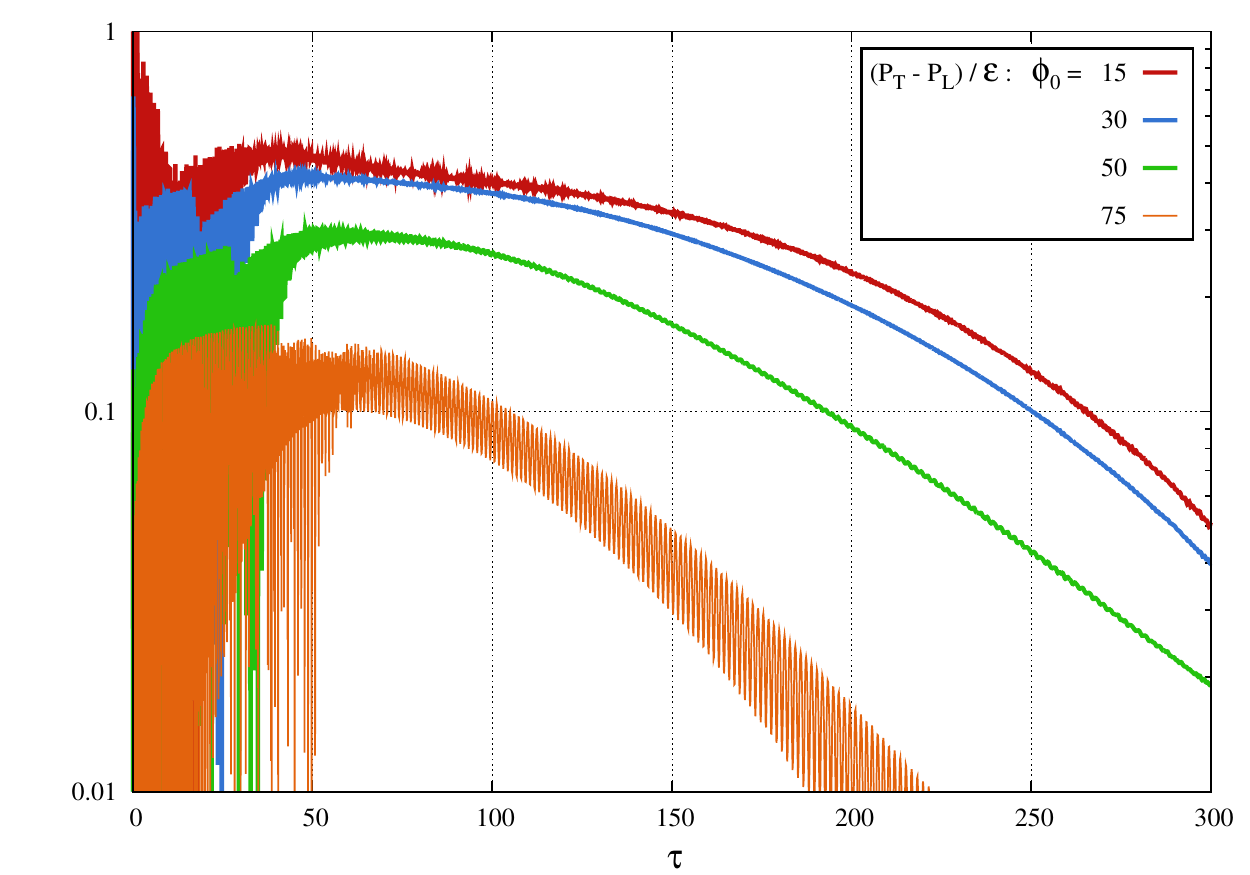}}
\end{center}
\caption{\label{fig:phi0dep} Dependence of the isotropization of the
  pressure on the amplitude of the background field. The coupling
  strength is the same in the four computations.}
\end{figure}
The scale invariance (at the classical level) of the theory we are
studying suggests that it should be generic for all background fields 
since a larger background field can be hidden in a rescaling of the
time and spatial coordinates.  Due to the limitations inherent to the
lattice setup, we can only vary the background field in a limited
window; nevertheless, we have indeed observed numerically that 
isotropization occurs regardless of the amplitude of the background
field. This is shown in the figure \ref{fig:phi0dep}, where the time
dependence of $(P_{_T}-P_{_L})/\epsilon$ is represented for uniform
background fields of amplitudes $\varphi_0=15,30,50,75$. Moreover, we
see that the isotropization time decreases if $\varphi_0$
increases. This decrease is expected because  all time scales
should be inversely proportional to the field amplitudes.  Note that
there are important lattice artifacts for the smallest value of
$\varphi_0$ that are visible in the very small change of the isotropization
curve when going from $\varphi_0=15$ to $\varphi_0=30$. In this
regime of smaller fields, the tadpole mass--due to the vacuum
fluctuations that we have not renormalized--seems to dominate the
dynamics. This is a lattice issue related to the breaking of the scale invariance by lattice cutoffs of the theory we are
simulating.

\begin{figure}[htbp]
\begin{center}
\resizebox*{10cm}{!}{\includegraphics{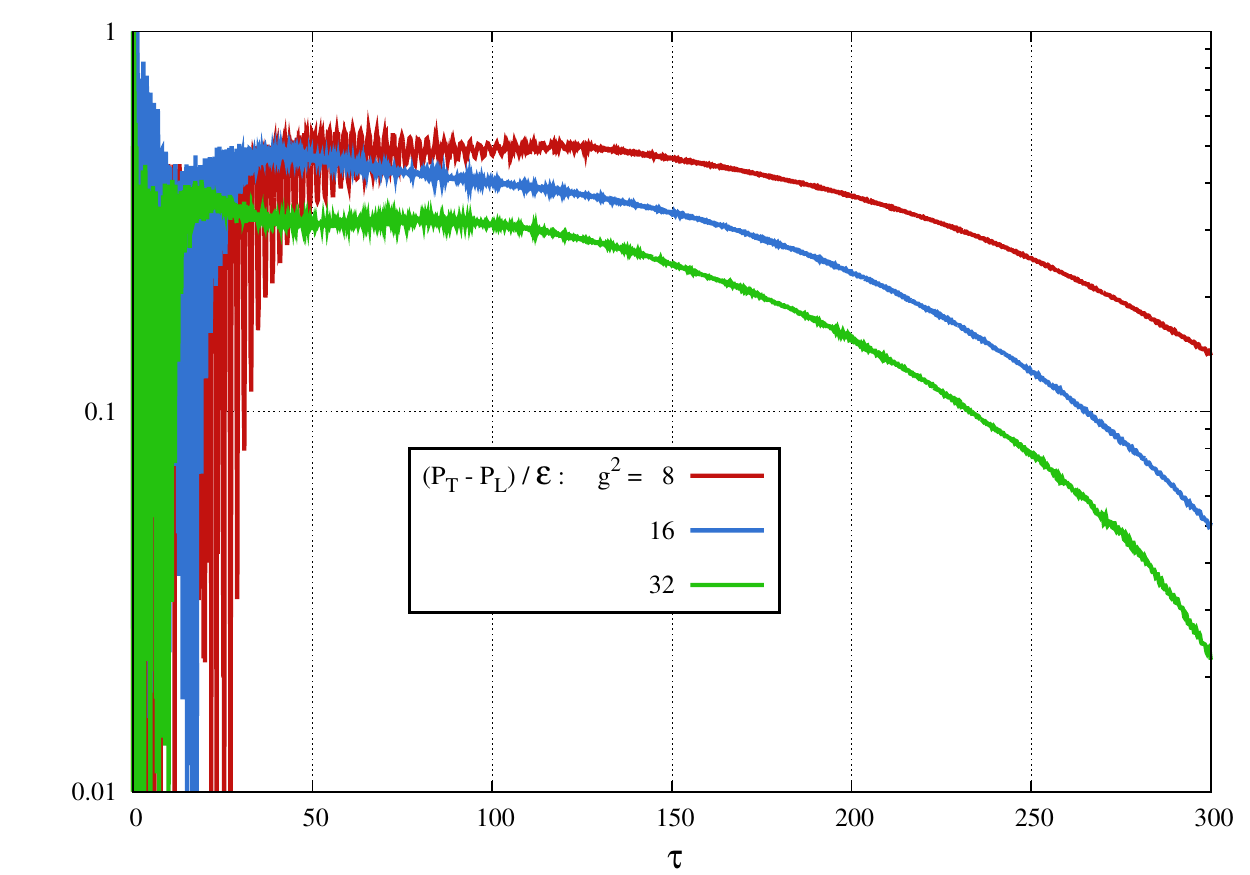}}
\end{center}
\caption{\label{fig:g2dep} Dependence of the isotropization of the
  pressure on the strength of the coupling. The amplitude of the
  background field is the same in the three computations.}
\end{figure}

Less obvious is the question of what happens when varying the
coupling. We do expect the isotropization time to decrease if
we increase the coupling; however,  there could be some critical coupling
below which isotropization never occurs. What we observe numerically over the  range of couplings explored (limited by the lattice configurations) is that isotropization is faster at larger coupling strengths,  as shown in fig. \ref{fig:g2dep}. 

\section{Computation of Hankel functions of imaginary index}
\label{app:hankel}
In the computation of the initial conditions (eq.~(\ref{eq:fluct1}))
and of the occupation number (eq.~(\ref{eq:fk})), we need Hankel
functions of imaginary index $H_{i\nu}^{(2)}$. They are both needed at
small arguments (for the initial conditions), and at fairly large
arguments (for the occupation number). And we need them for very large
indices $\nu$, since on a lattice with $N$ spacings in a unit rapidity
interval, the maximal value of $\nu$ is $\nu_{\rm max}=2N$. Thus, for
the value $N=320$ that we have used in most of this paper, we need
Hankel functions with indices up to $640$.

To the best of our knowledge, Hankel functions of imaginary index are
not implemented in the common scientific numerical libraries such as
CERNLIB or the GNU Scientific Library, but it is fairly easy to
evaluate them numerically with high accuracy. First of all, one should
recall that they are solutions of the Bessel equation
\begin{equation}
f^{\prime\prime}+\frac{1}{\tau}f^\prime+(1+\frac{\nu^2}{\tau^2})f=0\; .
\end{equation}
Naturally, a pair of linearly independent solutions of this equation
can be constructed numerically to any desired accuracy. But the
problem is to find the proper linear combination of those that gives
the desired Hankel function. It is best defined by its asymptotic
expansion,
\begin{eqnarray}
H_{i\nu}^{(2)}(\tau)
&=&
\left(\frac{2}{\pi\tau}\right)^{1/2}
e^{-i(\tau-\pi/4)}
e^{-\pi\nu/2}
\nonumber\\
&&\quad\times
\sum_{k=0}^n
\frac{(-1)^k}{k! (2i\tau)^k}
\,
\left(\nu^2+\frac{1^2}{4}\right)
\left(\nu^2+\frac{3^2}{4}\right)
\cdots
\left(\nu^2+\frac{(2k-1)^2}{4}\right)
+{\cal O}(\tau^{-n-1})\; .
\label{eq:Hasympt}
\end{eqnarray}
This expansion should be used with caution, because it is not a
convergent series if summed to arbitrarily large $n$. But at a given
$\nu$, one can find a sufficiently large $\tau$ and an optimal $n$
such that the residual term is extremely small. In practice, one
computes the successive terms in eq.~(\ref{eq:Hasympt}) for a given
$\nu$ and a large $\tau$: they decrease for small $k$, but
eventually start increasing again for $k$ larger than some $n$. The
last computed term before they start increasing is used as an estimate
of the error. If it is satisfactorily small, then we compute the
Hankel function and its first derivative by summing the asymptotic
expansion up to this $n$.  If the error is not small enough, we start
over this process for the same $\nu$, but a larger $\tau$.

At this point, we know the value of $H_{i\nu}^{(2)}$ (and that of its first
derivative) for some value of $\tau$. The next step is to use these as
initial conditions to solve numerically the Bessel equation, both
forward and backward in order to cover all the desired range of
arguments. This can be done to very high accuracy by using a high
order solver with adaptive steps.


\end{document}